\documentclass[american,prd,preprint,floatfix,letterpaper,showpacs,nofootinbib]{revtex4}
\usepackage[latin1]{inputenc}
\usepackage{amsmath}
\usepackage{amssymb}
\usepackage{color}
\usepackage{graphicx}

\makeatletter

\usepackage{wick}

\usepackage{babel}

\newcommand{\cM}{\ensuremath{\mathcal{M}}}

\newcommand{\cO}{\ensuremath{\mathcal{O}}}

\newcommand{\Str}{{\rm str}}

\newcommand*{\bea}{\begin{eqnarray}}
\newcommand*{\eea}{\end{eqnarray}}
\newcommand*{\be}{\begin{equation}}
\newcommand*{\ee}{\end{equation}}
\newcommand{\bra}{\langle}
\newcommand{\ket}{\rangle}

\newcommand*{\chpt}{$\chi$PT}

\makeatother

\begin{document}

\title{$K\to\pi$ and $K\to 0$ in 2+1 Flavor Partially Quenched Chiral Perturbation Theory}

\author{C.\ Aubin}
\email[]{caaubin@wm.edu}
\affiliation{Department of Physics, College of William and Mary, Williamsburg, VA 23187}
\author{J.\ Laiho}
\email[]{jlaiho@fnal.gov}
\affiliation{Physics Department, Washington University, St. Louis, MO 63130}
\author{S.\ Li}
\email[]{lishu@phys.columbia.edu}
\affiliation{Department of Physics, Columbia University, New York, NY 10027}
\author{M.F. Lin}
\email[]{meifeng@mit.edu}
\affiliation{Center for Theoretical Physics, Massachusetts Institute of Technology, Cambridge, MA 02139, USA}

\date{\today}

\begin{abstract}
We calculate results for $K\to\pi$ and $K\to 0$ matrix elements to next-to-leading order in 2+1 flavor partially quenched chiral perturbation theory.  Results are presented for both the $\Delta I=1/2$ and $3/2$ channels, for chiral operators corresponding to current-current, gluonic penguin, and electroweak penguin 4-quark operators.  These formulas are useful for studying the chiral behavior of currently available 2+1 flavor lattice QCD results, from which the low energy constants of the chiral effective theory can be determined.  The low energy constants of these matrix elements are necessary for an understanding of the $\Delta I=1/2$ rule, and for calculations of $\epsilon'/\epsilon$ using current lattice QCD simulations.
\end{abstract}

\maketitle

\section{Introduction}

Lattice QCD is a first principles approach to calculating low energy hadronic quantities using numerical Monte Carlo methods.  State-of-the-art calculations are now including 2+1 flavors of quarks in the weighting of the gauge configurations, thus eliminating the quenched approximation.  However, partially quenched simulations, where the valence quarks have different masses than those of the sea quarks, are still of use when combined with partially quenched chiral perturbation theory (PQ$\chi$PT) \cite{Bernard:1993ga}.  Since chiral perturbation theory ($\chi$PT) comes with a number of unknown low energy constants (LEC's), these LEC's must be obtained from non-perturbative methods, e.g., lattice calculations, or from experiment, in order to have predictive power.  When the number of light sea quarks is equal to three, then the LEC's of PQ$\chi$PT correspond to those of the unitary theory \cite{Sharpe:1997by, Sharpe:2000bc}, and the LEC's obtained from fits to partially quenched lattice data can be used to predict hadronic quantities.   Partial quenching can therefore be used in order to gain a better handle on chiral fits to numerical data, because varying the sea and valence quark masses separately leads to the determination of more linearly independent combinations of LEC's.  It also allows one to make use of more of the available lattice data, since simulating additional valence quark masses is relatively cheap compared to generating more ensembles with different sea quark masses. 

In this work we calculate PQ$\chi$PT expressions relevant for obtaining $K\to\pi\pi$ matrix elements from lattice simulations.
Although matrix elements of $K\to \pi\pi$ are of importance to phenomenology, there are difficulties with extracting multi-hadron decay amplitudes directly from the lattice, as expressed by the Maiani-Testa no-go theorem \cite{Maiani:1990ca}.  The implication of this no-go theorem is that physical amplitudes can only be computed if the final state pions are at rest, or some other unphysical set of kinematics.  It was shown by Lellouch and L\"uscher \cite{Lellouch:2000pv} (see also Ref.~\cite{Lin:2001ek}) that this no-go theorem can be evaded, and that the matrix elements can be computed at physical kinematics using finite volume correlation functions.  Although this method does not require $\chi$PT, the physical volume necessary to implement the method at physical quark masses is large, and therefore prohibitively expensive given the present computational resources.  

An alternative method for calculating $K\to\pi\pi$ from lattice QCD simulations is to obtain the leading order LEC's necessary to construct $K\to\pi\pi$ from lattice simulations of the simpler quantities $K\to\pi$ and $K\to 0$.  This method was introduced quite some time ago in Ref.~\cite{Bernard:1985wf}.  Given that there are large corrections to kaon matrix elements coming from chiral logarithms at higher orders in $SU(3)$ $\chi$PT, it is necessary to include next-to-leading order (NLO) corrections in the fits to lattice data.  This is true both because the light quark masses are still relatively heavy in present simulations, and also the physical strange quark mass is itself rather heavy.  It is an important, and as yet unanswered question whether the kaon mass is light enough so that $K\to\pi\pi$ amplitudes can be described by one-loop chiral perturbation theory to a useful precision. 
The issue of convergence is quantity dependent, and so must be studied for each quantity of interest.  We thus calculate the NLO PQ$\chi$PT expressions for $K\to\pi$ and $K\to 0$ matrix elements, including finite-size effects, which are needed both to extract LEC's from the lattice, and to assess the convergence of $\chi$PT by studying fits to lattice data as a function of quark masses.
	
In this work we calculate PQ$\chi$PT $K\to\pi$ and $K\to 0$ matrix elements in the isospin (2+1-flavor) limit.  We do not consider the completely non-degenerate quark mass case since isospin breaking leads to additional complications [such as (8,1)'s contributing to $\Delta I=3/2$ amplitudes], and these would also not be relevant to current lattice simulations.  Thus, we restrict ourselves to the 2+1 case in both the sea and valence sectors, but with no degeneracies between sea and valence quark masses.  We do not present here a complete set of formulas necessary to extract all of the NLO LEC's from 2+1 flavor lattice calculations, since some of the needed LEC's must be obtained from $K\to\pi\pi$ amplitudes at unphysical kinematics.  Even so, the formulas should be useful in extracting leading order LEC's from lattice data, and in studying the convergence of the chiral expansion.  Note that there are many works which discuss the determination of the LEC's needed to construct $K\to\pi\pi$ through NLO in $\chi$PT at physical kinematics \cite{Cirigliano:1999pv, Cirigliano:2001hs, Laiho:2002jq, Laiho:2003uy, Golterman:2000fw, Golterman:2001qj, Bijnens:1998mb, Lin:2002nq, Lin:2003tn, Kim:2007ri,Kim:2008dr}, though we make no attempt to review the various approaches here.

For the (8,1) ($\Delta I=1/2$) amplitudes there is an additional complication in the partially quenched theory coming from the treatment of the gluonic penguin 4-quark operator.  For the three-flavor theory the situation in PQ$\chi$PT is simplified significantly if the corresponding chiral operators are chosen to transform as (8,1)'s under the partially quenched graded symmetry group \cite{Golterman:2001qj, Golterman:2002us}.  That is the prescription we adopt in the current work.  If another choice is made, such as, for example, if the chiral operators are chosen to transform under the (8,1) chiral symmetry group of the full theory, then additional LEC's enter the calculation, making the determination of the desired LEC's more complicated.  Although this complication requires some care in the three flavor partially quenched theory, the method is still viable, unlike the quenched theory, in which quenched gluonic penguin amplitudes lead to large systematic uncertainties \cite{Golterman:2001qj,Golterman:2002us,Golterman:2003yw,Golterman:2003uj,Aubin:2006vt}.  

This paper is organized as follows: in Sec.~\ref{sec:pqchpt} we give a review of PQ$\chi$PT, including the effects of the weak Lagrangian, 
and in Sec.~\ref{sec:details} we give a quick overview of the calculation involved.
In Sec.~\ref{sec:subtractions} we review the operator subtraction that is necessary for $\Delta I=1/2$ matrix elements, and introduce the $\Theta^{(3,\overline{3})}$ operator for this purpose. NLO formulas of matrix elements of this operator are calculated for use in later sections. We present results for the (8,8) electroweak penguin operators for the $K\to 0$ and $K\to\pi$ processes in Sec.~\ref{sec:O88wme}, where we also give the physical $K\to\pi\pi$ amplitudes for completeness.  In Sec.~\ref{sec:O271wme32} we present results for the (27,1), $\Delta I=3/2$, $K\to\pi$ matrix element, and in Sec.~\ref{sec:O81_271_12} we present the results for (8,1) and (8,1)+(27,1) operators for $K\to 0$ and $K\to\pi$, including the operator subtraction. In Sec.~\ref{sec:FV} we discuss the finite volume corrections for the results presented in this work. We conclude in Sec.~\ref{sec:conc} and include relevant function definitions and the chiral logarithm contributions in a set of appendices.  Appendix~\ref{sub:erratum} provides an erratum for Refs.~\cite{Laiho:2002jq} and \cite{Laiho:2003uy}.
	
\section{Partially Quenched Chiral Perturbation Theory}\label{sec:pqchpt}

We use the standard formulation of partially quenched chiral perturbation theory (PQ\chpt) introduced in \cite{bernard:1992mk,bernard:1993sv}.
In this formulation, the valence quark loops are removed by
introducing ``ghost'' quarks with the same masses and
quantum numbers as their valence counterparts, but which obey opposite statistics. The chiral symmetry group for a partially quenched theory is graded; in general one takes it to be $SU\left(N_{\rm val}+N_{\rm sea}\,|\,N_{\rm val}\right)_{L}\otimes SU\left(N_{\rm val}+N_{\rm sea}\,|\,N_{\rm val}\right)_{R}$. For the purposes of this work, we set $N_{\rm val} = N_{\rm sea} =3$.
Specifically, we have three valence quarks denoted as $x$, $y$, and $z$; three sea quarks denoted as
$u$, $d$, and $s$; and finally three ghosts: $\tilde{x}$,
$\tilde{y}$, and $\tilde{z}$. 

\subsection{Strong Lagrangian in PQ\chpt }

As explained in Ref. \cite{Laiho:2003uy}, in the partially quenched
theory, operators are written in terms of the chiral field 
\begin{equation}
	\Sigma = \exp\left[\frac{2i\Phi}{f}\right]\ ,
	\label{eq:Sigma}
\end{equation}
where $f$ is the meson decay constant in the $SU(3)$ chiral limit (normalized such that the physical $f_\pi\approx 130.7$ MeV), $\Phi$ is a $9\times9$ matrix containing the meson fields,
\begin{equation}
	\Phi\equiv\left(\begin{array}{cc}
	\phi & \chi^{\dagger}\\
	\chi & \tilde{\phi}\end{array}\right)\ ,
	\label{eq:Phi}
\end{equation}
where $\phi$ is a $6\times6$ matrix of pseudoscalar mesons
constructed out of valence and sea quarks, $\tilde{\phi}$ is a
$3\times3$ matrix containing mesons constructed with two ghost
quarks, $\chi$ ($\chi^{\dagger}$) is a $3\times6$ ($6\times3$) matrix containing fermionic mesons made out of one quark and one ghost quark. $\Sigma$ transforms under the graded chiral symmetry group as
\begin{equation}
	\Sigma \to L\Sigma R^\dag\ ,
\end{equation}
with $L\in SU\left(6\,|\,3\right)_{L}, R\in SU\left(6\,|\,3\right)_{R}$.  Operators in the chiral effective theory are constructed from the quark-level operators out of $\Sigma$ and other objects (such as the quark charge matrix and mass matrix, for example) such that they transform the same way under the chiral symmetry group. 

The leading-order (LO) strong Lagrangian is given by \cite{Gasser:1984gg}
\begin{equation}
	\mathcal{L}_{st}^{\left(2\right)}=
	\frac{f^{2}}{8}\mathrm{str}
	\left[\partial_{\mu}\Sigma\partial^{\mu}\Sigma^{\dagger}\right]
	+\frac{f^{2}B_{0}}{4}\mathrm{str}\left[ \cM \Sigma 
	+ \Sigma^{\dagger} \cM^\dag \right],
	\label{eq:strong lagrangian p2}
\end{equation}
where the superscript (2) indicates that this Lagrangian is valid to $\mathcal{O}\left(p^{2}\right)$ in the chiral power counting scheme, 
and $\cM$ is the quark mass matrix
\begin{equation}
	\cM = \mathrm{diag}\left(m_{x},m_{y},m_{z},m_{u},
	m_{d},m_{s},m_{x},m_{y},m_{z}\right)\ .
	\label{eq:quark mass}
\end{equation}
Note that this corresponds to the quark vector composed of valence quarks, sea quarks, and ghost quarks
\begin{equation}
	q=\left(x,y,z,u,d,s,\tilde{x},\tilde{y},
	\tilde{z}\right)^{\mathrm{T}}.
	\label{eq:quark vector}
\end{equation}
The supertrace is defined as follows: for a $9\times9$ matrix
\begin{equation}
	U_{9\times9}=\left(\begin{array}{cc}
	A_{6\times6} & B_{6\times3}\\
	C_{3\times6} & D_{3\times3}\end{array}\right)\label{eq:U9x9}
\end{equation}
in which sub-matrix $A$ is the top-left $6\times6$ diagonal block
and $D$ is the bottom-right $3\times3$ diagonal block, then 
\begin{equation}
	\mathrm{str}\left(U\right)=\mathrm{tr}\left(A\right)
	-\mathrm{tr}\left(D\right).\label{eq:supertrace}
\end{equation}

We set the valence $x$ and $y$ quark masses equal, and we set the sea $u$ and $d$ quark masses equal,
\begin{equation}
	m_{x}  =  m_{y},\quad	 m_{u}  =  m_{d}.
\label{eq:mu = md}
\end{equation}
Thus we work in the isospin limit in both the valence and sea sector, and we present results for both this (2+1-flavor) case and the 3-
flavor case (degenerate valence quarks).

At NLO in the full theory [$\mathcal{O}\left(p^{4}\right)$], the
strong Lagrangian involves 12 additional operators with undetermined
coefficients \cite{Gasser:1983yg, Gasser:1984gg}.  There is an additional $\mathcal{O}(p^4)$ operator which appears in the partially quenched theory \cite{Sharpe:2003vy}, though this operator does not contribute to the quantities considered in this work.  The 
NLO operators of the strong Lagrangian relevant for the current work are
\begin{align}
 \mathcal{O}^{(st)}_4 &= \mathrm{str}\left[ L^2 \right] \mathrm{str}\left[ S \right], \nonumber\\  
 \mathcal{O}^{(st)}_5 &= \mathrm{str}\left[ L^2 S \right], \nonumber\\   
 \mathcal{O}^{(st)}_6 &= \mathrm{str}\left[ S \right]^2, \nonumber\\   
 \mathcal{O}^{(st)}_8 &= \frac{1}{2}\mathrm{str}\left[ S^2 - P^2 \right] \ , \label{eq: strong op p4}     
\end{align}
where \begin{align}
S & =2B_{0}\left(\cM^{\dagger}\Sigma^{\dagger}+\Sigma \cM \right),\nonumber \\
P & =2B_{0}\left(\cM^{\dagger}\Sigma^{\dagger}-\Sigma \cM \right),\nonumber \\
L_{\mu} & =i\Sigma \partial_{\mu} \Sigma^{\dagger}.             \label{eq:NLO_pieces}
\end{align}

As follows from the strong Lagrangian above, the leading-order
mass of a bare pseudo-scalar meson is 
\begin{equation}
	m_{ij}^{2}=B_{0}\left(m_{i}+m_{j}\right)\ ,
	\label{eq:meson mass}
\end{equation}
where $m_{ij}$ is the mass of meson $\Phi_{ij}$, $m_{i}$ and $m_{j}$
are the masses of the quarks $q_{i}$ and $q_{j}$ ($i,j$ can refer to the sea, valence, or ghost quarks in this case). In our partially quenched amplitudes we assume that the light quark masses are all light enough compared to the $\eta'$ mass so that the $\eta'$ can be integrated out.
As demonstrated in Ref.~\cite{Sharpe:2000bc}, this is the case where
the LEC's of the partially quenched theory with three sea quarks correspond to the LEC's of the unitary theory.

In the following we adopt the notation that the masses of mesons which are constructed out of two different flavors of quarks are labelled in terms of their quark consituents, regardless of whether they are sea or valence, for example $m_{xy}$ or $m_{zs}$. For any flavor-neutral meson, we use $m_D \equiv m_{dd}$ or $m_X\equiv m_{xx}$ for mesons in the ``flavor basis.'' Due to the disconnected propagators which arise in the flavor-neutral sector, this is distinct from the ``physical basis,'' where the relevant mesons are the $\pi^0$ and $\eta$. 
These only arise in the sea sector, and we will use the fact that
\begin{eqnarray}
	m_{\pi^0}^2 & = & m_D^2\ , \nonumber\\
	m_\eta^2 & = & \frac{1}{3}\left( 2m_S^2 + m_D^2\right),\nonumber
\end{eqnarray}
in the isospin limit ($m_u = m_d$).

The propagators for flavor neutral mesons are obtained by following
the prescription in Ref.\cite{Sharpe:2000bc} in Minkowski space:
\begin{align}
	\wick{1}{<1\Phi_{ii}>1\Phi_{jj}} & =
	\frac{i\delta_{ij}\varepsilon_{i}}{p^{2}-m_{ii}^{2}
	+i\epsilon}-\frac{i}{3}\frac{\left(p^{2}-m_{D}^{2}\right)
	\left(p^{2}-m_{S}^{2}\right)}{\left(p^{2}-m_{ii}^{2}\right)
	\left(p^{2}-m_{jj}^{2}\right)\left(p^{2}-m_{\eta}^{2}\right)}\ .
	\label{eq:prop neutral}
\end{align}
The propagators for flavor off-diagonal mesons are
\begin{equation}
	\wick{1}{<1\Phi_{ij}>1\Phi_{ji}}=\frac{i\varepsilon_{j}}
	{p^{2}-m_{ij}^{2}+i\epsilon}\label{eq:prop off-diag}
\end{equation}
where
\begin{equation}
	\varepsilon_{j}=\begin{cases}
	1 & j\in\{ x,y,z,u,d,s\}\\
	-1 & j\in\{\tilde{x},\tilde{y},\tilde{z}\}\end{cases}\ .
	\label{eq:propsign}
\end{equation}

\subsection{Leading-Order Weak Lagrangian\label{sub:Quark-Charges}}

In full \chpt, we group the weak operators appearing in the
$K\to\pi\pi$ transition by their chiral transformation properties
in the $SU\left(3\right)_L\otimes SU\left(3\right)_R$ symmetry group. We can carry this same idea over to 
PQ\chpt, where we extend their definition into the graded group
using the expanded chiral field $\Sigma$ and replacing traces with
supertraces. Except for those cases discussed explicitly (such as the quark charge and mass matrices, for example), when going from unquenched to partially quenched $\chi$PT, operators are replaced as follows:
\begin{equation}
	\lambda \to \left(
	\begin{matrix}\lambda_{3\times3} & 0_{3\times6} \\
	 0_{6\times3} & 0_{6\times6}\end{matrix}
	\right)\ ,
\end{equation}
where the upper left block of this matrix is the $3\times3$ block corrsponding to the valence sector.

The leading order weak operators are \cite{Bernard:1985wf,Cirigliano:1999pv,Cirigliano:2001hs,Laiho:2003uy,Blum:2001xb}\begin{align}
	\mathcal{O}_{LO}^{\left(8,8\right)} & 
    =\mathrm{str}\left[\lambda_{6}\Sigma Q\Sigma^{\dagger}\right]     
	\nonumber \\
	\mathcal{O}_{LO,1}^{\left(8,1\right)} &
    =\mathrm{str}\left[\lambda_{6}\partial_{\mu}
    \Sigma\partial^{\mu}\Sigma^{\dagger}\right]    
	\nonumber \\
	\mathcal{O}_{LO,2}^{\left(8,1\right)} &
     =2B_{0}\mathrm{str}\left[\lambda_{6}\left(\Sigma \cM+\cM^{\dagger}	
     \Sigma^{\dagger}\right)\right]
	\nonumber \\
	\mathcal{O}_{LO}^{\left(27,1\right)} &
    =t_{kl}^{ij}\left(\Sigma\partial_{\mu}\Sigma^{\dagger}
    \right)_{i}^{k}
	\left(\Sigma\partial^{\mu}\Sigma^{\dagger}\right)_{j}^{l}
	\label{eq:weak operator p2}
\end{align}
where $Q$ is the quark charge matrix, $(\lambda_{6})_{ij}=\delta_{i3}\delta_{j2}$, and the tensor $t_{kl}^{ij}$ is symmetric on any indices and traceless on pairs of upper and lower indices, and its elements are chosen to
pick out the $\Delta S=1$ transitions; it thus plays a similar role to that of $\lambda_{6}$
for the other operators. However, we will defer the actual determination
of its non-zero elements until Section \ref{sub:O27def}, where we will use this
tensor to further divide the operator into the isospin 3/2 part and
the isospin 1/2 part and directly evaluate their respective amplitudes.  The isospin decomposition of $K\to\pi$ matrix elements is given in Appendix~\ref{sec:Isospin-Decomposition}.

There is a choice to be made for the quark charge matrix, $Q$, above, which enters in the electroweak penguin operators
\cite{Laiho:2003uy}. We could either assign charges to ghosts such
that they cancel out the electroweak valence quark loops, or we could
make them uncharged. In this paper we derive the amplitudes with the
electroweak penguin operators for both choices, which we denote as
$Q_{1}$ and $Q_{2}$.  We
always assign zero charge to the sea quarks, since that is what is typically done when generating lattice gauge fields. The two
choices of charge matrix are:
\begin{align}
	Q_{1} & =\mathrm{diag}\left(2,-1,-1,0,0,0,2,-1,-1
	\right),\nonumber \\
	Q_{2} & =\mathrm{diag}\left(2,-1,-1,0,0,0,0,0,0\right).
	\label{eq:quark charges}
\end{align}

We discuss the weak operators which contribute to $K\to\pi$
and $K\to\pi\pi$ at next-to-leading order in subsequent sections.

\section{Details of the calculation}\label{sec:details}

To make complete use of lattice data in extracting LEC's relevant for $K\to\pi\pi$, it is important to work in the non-degenerate $m_x=m_y\ne m_z$ case.  Since the $K\to\pi$ amplitudes do not conserve 4-momentum for $m_z\ne m_x$, the weak operator must transfer a 4 momentum $q\equiv p_{xz}-p_X$.  In our calculations we restrict ourselves to the case where both initial and final mesons are at rest, so $q=(m_{xz}-m_X, 0, 0, 0)$.

\begin{figure}
\begin{center}
\includegraphics[scale=.8]{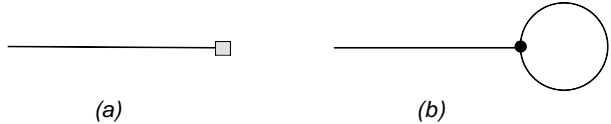}
\caption{Diagrams contributing to $K\to 0$ at NLO.  The gray square is the insertion of a NLO weak vertex, and the small dot is an insertion of the LO weak vertex. \label{fig:k2vac}}
\end{center}
\end{figure}

\begin{figure}
\begin{center}
\includegraphics[scale=.8]{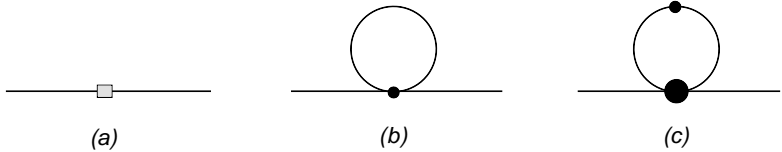}
\caption{Diagrams contributing to $K\to\pi$ at NLO.  As in Fig.~\ref{fig:k2vac}, the gray square is an insertion of a NLO weak vertex, and the small dot is an insertion of the LO weak vertex.  The large dot is the insertion of an $O(p^2)$ strong vertex. \label{fig:k2pi}}
\end{center}
\end{figure}

The NLO diagrams contributing to $K\to 0$ and $K\to\pi$ are given in Figs.~\ref{fig:k2vac} and \ref{fig:k2pi}, respectively.  The external legs are always mesons made of two valence quarks, while the internal loops in the partially-quenched theory consist of valence-ghost, valence-sea, and valence-valence mesons.  In addition to these diagrams, the renormalization of the external legs (wave-function renormalization) via the strong interactions must be taken into account.

The logarithmic expressions presented in the appendices of this work are quite lengthy.  Thus, checks are necessary.  The first check was that the one-loop insertions cancel those of the divergent counterterms, and this check was performed for all expressions in this paper.  Another check is that an expression reduces to some other in the appropriate limit.  All of the logarithmic expressions in this paper reduce to those in Refs.~\cite{Cirigliano:1999pv,Golterman:2000fw,Cirigliano:2001hs,Laiho:2002jq,Laiho:2003uy} in the appropriate degenerate sea quark and full QCD limits.\footnote{Note that Ref.~\cite{Cirigliano:1999pv} contains errors that are corrected in Ref.~\cite{Cirigliano:2001hs}, and we agree with the latter.}  Finally, all one-loop expressions in this work were computed separately by at least two of the authors, using independently written code. The codes used were the FEYNCALC package \cite{Mertig:1990an} written for the Mathematica \cite{Wolfram:1988} system, and the FORMCalc package \cite{Hahn:1998yk}, which interfaces FORM \cite{Vermaseren:2000nd} with Mathematica.

\section{Subtraction of $\Delta I=1/2$ Amplitudes}\label{sec:subtractions}

In general the $\Delta I=1/2$ matrix elements of four-quark operators have a power divergent part due to the four-quark operators mixing under renormalization with lower dimensional operators when using a lattice regularization.  This power divergence reduces to a quark bilinear times a momentum independent coefficient \cite{Blum:2001xb}.
Following Refs. \cite{Bernard:1985wf, Blum:2001xb, Laiho:2003uy}, in order to remove
the power-divergence of $\Delta I=1/2$ operators, we perform
a subtraction using the dimension three quark-level operator,
\begin{equation}
	\Theta^{\left(3,\bar{3}\right)}\equiv\bar{s}
	\left(1-\gamma_{5}\right)d\ .
	\label{eq:theta-3,3}
\end{equation}
This subtraction must also be performed in PQ$\chi$PT for comparison with the subtracted lattice results.
Again following Refs.~\cite{Bernard:1985wf,Blum:2001xb,Laiho:2003uy}, the lowest order [$\cO(p^0)$] chiral operator corresponding to the $(3,\bar3)$ operator in Eq.~(\ref{eq:theta-3,3}) is
\begin{align}
	\Theta^{\left(3,\bar{3}\right)}_{LO} = 
	& \alpha^{\left(3,\bar{3}\right)} 
	\mathrm{str}\left[\lambda_6 \Sigma\right] \ ,
\end{align}
where the low-energy constant $\alpha^{\left(3,\bar{3}\right)}$ can be related to the coefficient of the mass term in the leading-order strong Lagrangian \cite{Laiho:2003uy},
\[
	\alpha^{\left(3,\bar{3}\right)} = - \frac{f^2 B_0}{2}\ .
\]

As explained in Ref \cite{Blum:2001xb}, the $\Theta^{\left(3,\bar{3}\right)}$ operator can be used to remove the power divergences to all orders in the lattice calculation.  This subtraction is performed to NLO in PQ$\chi$PT explicitly in the sections that follow.
To this end, we require the higher order chiral operators of $\Theta^{\left(3,\bar{3}\right)}$.  The terms up to $O(p^2)$ needed for this work are
\begin{equation}
	\Theta^{\left(3,\bar{3}\right)}
	=
	\Theta^{\left(3,\bar{3}\right)}_{LO}
	+
	\sum_i c_{33,i}\cO'_i
\end{equation}
where $i$ takes the values $4,5,6,8, H_2$, and where
\begin{eqnarray}
 	\mathcal{O}'_4 & = &
	\frac{1}{2} \mathrm{str}\left[\lambda_6 \Sigma\right] 
 	\mathrm{str}\left[\partial_\mu\Sigma^\dagger
	\partial^\mu \Sigma\right], \nonumber \\
	\mathcal{O}'_5 & = &
	\frac{1}{2} \mathrm{str}\left[\lambda_6 \Sigma
	\partial_\mu\Sigma^\dagger
	\partial^\mu \Sigma\right], \nonumber \\
 	\mathcal{O}'_6 & = & 
	2 B_0\mathrm{str}\left[\lambda_6 \Sigma\right] 
    \mathrm{str}\left[ \cM^\dagger \Sigma 
    + \Sigma^\dagger \cM\right], \nonumber \\
 	\mathcal{O}'_8 & = &
	2 B_0 \mathrm{str}\left[\lambda_6 \Sigma 
	\cM^\dagger \Sigma\right], \nonumber \\
 	\mathcal{O}'_{H_2} & = & 
	B_0\mathrm{str}\left[\lambda_6 \cM\right].  
 \label{eq:NLO op 3,3}
\end{eqnarray}
The coefficients $c_{33,i}$ of the operators $\mathcal{O}'_i$ are related to the Gasser-Leutwyler coefficients by $c_{33,i} = -8 B_0 L_i$, a relation similar to that for the leading order coefficient, $\alpha^{\left(3,\bar{3}\right)}$, given above.

To NLO, the $K\to 0$ matrix element for 2+1 valence flavors is
\begin{eqnarray}
	\left\langle 0\left|\Theta^{\left(3,\bar{3}\right)}
	\right|K^{0}\right\rangle
	& = & \frac{2i}{f}\alpha^{\left(3,\bar{3}\right)}
	\left[1+\frac{1}{2}\delta Z_{xz}\right]
	\nonumber \\&&{}
	+ \frac{4i}{9}\frac{\alpha^{\left(3,\bar{3}\right)}}{f^{3}}
	\biggl\{\biggl[1+R_{X}(m_{\eta},m_{Z})-R_{\eta}(m_{X},m_{X})
	\biggr]\ell(m_{X}^{2})
	\nonumber \\&&{}
 	+\biggl[1+R_{Z}(m_{\eta},m_{X})-R_{\eta}(m_{Z},m_{Z})\biggr]
	\ell(m_{Z}^{2})
	\nonumber \\&&{}
	-6\ell(m_{xd}^{2})-3\ell(m_{xs}^{2})-6\ell(m_{zd}^{2})
	-3\ell(m_{zs}^{2})
	\nonumber \\&&{}
 	+\biggl[R_{\eta}(m_{X},m_{X})+R_{\eta}(m_{X},m_{Z})
	+R_{\eta}(m_{Z},m_{Z})\biggr]\ell(m_{\eta}^{2})
	\nonumber \\&&{}
	-R_{X}(m_{\eta})\tilde{\ell}(m_{X}^{2})-R_{Z}(m_{\eta})
	\tilde{\ell}(m_{Z}^{2})\biggr\} 
	\nonumber \\&&{}
	- \frac{32 i B_0}{f}\biggl\{L_8 m_{xz}^{2} 
	+ L_6 \left(2m_D^2+m_S^2\right) \biggr\}, \label{eq:theta-3,3 K->0}
\end{eqnarray}
where the chiral logarithms $\ell(m^2)$ and $\tilde{\ell}(m^2)$ are defined in Appendix~\ref{sec:loopfunc}, along with the residues $R_x(m_a)$, $R_x(m_a, m_b)$.  The wave-function renormalization $\delta Z_{xz}$ is given in Appendix~\ref{sec:renorms}. 

To NLO, the $K\to\pi$ matrix element (also for 2+1 valence flavors) is
\begin{eqnarray}
 	\left\langle \pi^{+}\left|\Theta^{\left(3,\bar{3}\right)}
	\right|K^{+}\right\rangle 
	& = & -\frac{2}{f^{2}}\alpha^{\left(3,\bar{3}\right)}
 	- \frac{16B_0}{f^{2}} \biggl\{ L_5 m_{X} m_{xz} - 2L_8 
	\left( m_{X}^{2} + m_{xz}^{2} \right)
	\nonumber \\&&{}
	 - 2L_6 \left( 2m_D^2+m_S^2 \right) \biggr\} 
	+ \left\langle \pi^{+}\left|\Theta^{\left(3,\bar{3}\right)}
	\right|K^{+}\right\rangle_{\rm logs}
	\label{eq:theta-3,3 K->pi ct}
\end{eqnarray}
For clarity, the rather lengthy logarithmic contribution is given in Appendix~\ref{sub:logterms 33}.
For degenerate valence masses ($m_x = m_y = m_z$), the $K\to\pi$ amplitude simplifies to
\begin{align}
 	\left\langle \pi^{+}\left|\Theta^{(3,\bar{3})}
	 \right|K^{+}\right\rangle ^{\rm deg.val.}
	 = & -\frac{2}{f^{2}}\alpha^{(3,\bar{3})}
	\left[1+\delta Z_{X}\right]
	\nonumber \\
	& + \frac{4}{3}\frac{\alpha^{(3,\bar{3})}}{f ^{4} }
	 \biggl\{m_{X}^{2}\biggl[R_{X}(m_{\eta})\tilde{\tilde{\ell}}(m_{X}^{2})
	 -2R_{\eta}(m_{X},m_{X})\beta(0,m_{\eta}^{2},m_{X}^{2})\biggr] \nonumber\\
 	&  +\biggl[-1  +R_{\eta}(m_{X},m_{X})\biggr]\ell(m_{X}^{2})  
	+2\ell(m_{xd}^{2})  +\ell(m_{xs}^{2})  \nonumber\\
	& -R_{\eta}(m_{X},m_{X})\ell(m_{\eta}^{2}) \nonumber\\
 	&  +\biggl[2m_{X}^{2}\biggl(1  -R_{\eta}(m_{X},m_{X})\biggr)
	+R_{X}(m_{\eta})\biggr]\tilde{\ell}(m_{X}^{2})\biggr\}  \nonumber\\
	&  - \frac{16B_0}{f^{2}} \biggl\{ \left( L_5 - 4L_8 \right) 
	m_{X}^{2} - 2L_6 \left( 2m_D^2+m_S^2 \right) \biggr\} 
	\label{eq:theta-3,3 K->pi deg val}, 
\end{align}
where $\tilde{\tilde{\ell}}(m^2)$ and $\beta(q^2, m^2_1, m^2_2)$ are defined in Appendix~\ref{sec:loopfunc}, and $\delta Z_X$ is given in Appendix~~\ref{sec:renorms}. These expressions are used below when performing the power divergent operator subtractions that are necessary in order to obtain the physical amplitudes in which we are interested.

\section{Weak Matrix Elements with $(8,8)$, $\Delta I=3/2$ and $1/2$ operators }\label{sec:O88wme}

In this section we present the results for the chiral operators which transform as (8,8)'s under the chiral symmetry.  These correspond to the electroweak penguin 4-quark operators.  Formulas are presented for $K\to 0$ and $K\to \pi$ for nondegenerate ($m_x=m_y\ne m_z$) valence quark masses, as well as for $K\to \pi$ with degenerate valence quark masses.  The power divergent subtraction is discussed for the $\Delta I=1/2$, $K\to\pi$ amplitude.  Since $K\to 0$ and $K\to\pi$ are sufficient to construct $K\to\pi\pi$ to NLO at physical kinematics for the (8,8)'s, we present the physical $K\to\pi\pi$ amplitudes as well.  

Following Ref.~\cite{Laiho:2003uy}, the form of the 
operator $\mathcal{O}^{\left(8,8\right)}$ through NLO
in PQ\chpt\ is
\begin{align}
\mathcal{O}^{\left(8,8\right)}= & 
	\alpha_{88}\mathrm{str}\left[\lambda_{6}\Sigma Q \Sigma^{\dagger}\right]              \nonumber \\
	& +c_{88,1}\mathrm{str}\left[\lambda_{6}L_{\mu} \Sigma Q \Sigma^{\dagger} L^{\mu} \right] \nonumber \\
	& +c_{88,2}\mathrm{str}\left[\lambda_{6}L_{\mu}\right]
			\mathrm{str}\left[\Sigma Q \Sigma^{\dagger} L^{\mu} \right]    \nonumber \\
	& +c_{88,3}\mathrm{str}\left[\lambda_{6}
			\left\{ \Sigma Q \Sigma^{\dagger}, L^{2} \right\} \right]      \nonumber \\
	& +c_{88,4}\mathrm{str}\left[\lambda_{6}
			\left\{ \Sigma Q \Sigma^{\dagger}, S \right\} \right]          \nonumber \\
	& +c_{88,5}\mathrm{str}\left[\lambda_{6}
			\left[\Sigma Q \Sigma^{\dagger}, P \right] \right]             \nonumber \\
	& +c_{88,6}\mathrm{str}\left[\lambda_{6} \Sigma Q \Sigma^{\dagger} \right]
			\mathrm{str}\left[ S \right]\ ,
			\label{eq:O88}
\end{align}
with $S,P,$ and $L_\mu$ as defined in Eq.~(\ref{eq:NLO_pieces}). 

\subsection{$K\to 0$ amplitudes for 2+1 valence flavors}

As explained in Section \ref{sub:Quark-Charges}, there are two choices
for the quark charge matrix for the operators in the (8,8) representation. If we set $Q=Q_1$, we obtain for the $K\to 0$ amplitude
\begin{eqnarray}
 	\left\langle 0\left|\mathcal{O}^{\left(8,8\right)}
	\right|K^{0}\right\rangle _{Q_{1}} 
	& = & 
	\frac{4i}{f^{3}}\alpha_{88}\left[   
    -2\ell\left(m_{xd}^{2}\right)-\ell\left(m_{xs}^{2}\right)
    +2\ell\left(m_{zd}^{2}\right)+\ell\left(m_{zs}^{2}\right)
    \right]
	\nonumber\\
	&&{}-\frac{8i}{f}c_{88,4}\left(m_{xz}^{2}-m_{X}^{2}\right)    
	\ .
	\label{eq:O88 K->0 Q1}
\end{eqnarray}

If we set $Q = Q_{2}$, we obtain
\begin{eqnarray}
 	\left\langle 0\left|\mathcal{O}^{\left(8,8\right)}
	\right|K^{0}\right\rangle _{Q_{2}}
	& = & \frac{4i}{f^{3}}\alpha_{88}\bigl\{ 
        -\ell\left(m_{X}^{2}\right)
        +2\ell\left(m_{xz}^{2}\right)
	-\ell\left(m_{Z}^{2}\right)
	-2\ell\left(m_{xd}^{2}\right)
	-\ell\left(m_{xs}^{2}\right)
	\nonumber \\
 && {}+2\ell\left(m_{zd}^{2}\right)
	+\ell\left(m_{zs}^{2}\right)   
	 \bigr\} 
	 -\frac{8i}{f}c_{88,4}\left(m_{xz}^{2}-m_{X}^{2}\right)      \label{eq:O88 K->0 Q2}\ .
 \end{eqnarray}

\subsection{$K\to\pi$ amplitudes for 2+1 valence flavors}

The process $K\to\pi$ must be separated into its $\Delta I=3/2$ and $\Delta I=1/2$ pieces, and we give the explicit isospin decomposition in Appendix \ref{sec:Isospin-Decomposition}. For the $\Delta I=3/2$ amplitudes we have 
\begin{eqnarray}
	\left\langle \pi^{+}\left|\mathcal{O}^{\left(8,8\right)(3/2)}
	\right|K^{+}\right\rangle _{Q_{2}}
	& = &
	\left\langle \pi^{+}\left|\mathcal{O}^{\left(8,8\right)(3/2)}
	\right|K^{+}\right\rangle _{Q_{1}}\nonumber \\
	& = & \frac{4\alpha_{88}}{f^{2}}
	+ \frac{4}{f^{2}}\biggl\{
    -\left(c_{88,1}+c_{88,2}\right)m_{xz}m_{X}
    \nonumber \\&& {}
    +2\left(c_{88,4}+c_{88,5}\right)\left(m_{xz}^{2}+m_{X}^{2}\right)
	+2c_{88,6}\left(2m_D^2+m_S^2\right)
	\biggr\}
    \nonumber \\&& {}
	+ \left\langle \pi^{+}
	\left|\mathcal{O}^{\left(8,8\right)(3/2)}
	\right|K^{+}\right\rangle_{Q_{1},{\rm logs}}\ .
	\label{eq:O88 Kpi Q1 3/2 ct}
\end{eqnarray}
For brevity, we
give only the analytic part of these matrix elements here; the logarithmic contributions are given in Appendix~\ref{sub:logterms 88}.

For the $\Delta I = 1/2$ amplitudes, we are ultimately interested in the subtracted versions, as discussed in Sec.~\ref{sec:subtractions}. 
We expand the amplitude $\left\langle 0\left|\Theta^{\left(3,\bar{3}\right)}\right|K^{0}\right\rangle $
to leading nontrivial order, and take the ratio
\begin{align}
	\frac{\left\langle 0\left|\mathcal{O}^{\left(8,8\right)}\right|K^{0}\right\rangle }
	{\left\langle 0\left|\Theta^{\left(3,\bar{3}\right)}
	\right|K^{0}\right\rangle } & =-\frac{4c_{88,4}}
	{\alpha^{\left(3,\bar{3}\right)}}
	B_0\left(m_{z}-m_{x}\right)+\frac{2}{f^{2}}
	\frac{\alpha_{88}}{\alpha^{\left(3,\bar{3}\right)}}
	\left({\rm logs}\right)+\dots\label{eq:O88 K->0 ratio}
\end{align}
where higher order terms in chiral perturbation theory are omitted.  The power divergent contribution is proportional to $m_z-m_x$, and this is true to all orders in the chiral expansion by CPS symmetry \cite{Bernard:1987pr}.
Thus, the ratio of LEC's containing the power divergence, $-4c_{88,4}B_0/(\alpha^{\left(3,\bar{3}\right)})$, can be extracted from the corresponding lattice matrix elements, since the mass dependence of the divergent piece is known to all orders of the chiral expansion.
We perform the operator subtraction using this ratio
and the amplitude 
$\left\langle \pi^{+}\left|\Theta^{\left(3,\bar{3}\right)}\right|K^{+}\right\rangle $,
\bea
 	\left\langle \pi^{+}\left|
	\mathcal{O}^{(8,8)(1/2)}_{\rm sub}
	\right|K^{+}\right\rangle_{Q}
	&=&  \left\langle \pi^{+}\left|
	\mathcal{O}^{(8,8)(1/2)}
	\right|K^{+}\right\rangle_Q
	\nonumber \\ && +\frac{4c_{88,4}B_0(m_z+m_x)}{\alpha^{(3,\bar{3})}}
	\left\langle \pi^{+}\left|\Theta^{\left(3,\bar{3}\right)}
	\right|K^{+}\right\rangle,
	\label{eq:O88 K->pi sub1}
\eea
where by CPS symmetry the power divergence is removed to all orders in $\chi$PT.  Through NLO in $\chi$PT we have, 
\bea
 	 \left\langle \pi^{+}\left|
	\mathcal{O}^{(8,8)(1/2)}_{\rm sub}
	\right|K^{+}\right\rangle_{Q}
	=  \left\langle \pi^{+}\left|
	\mathcal{O}^{\left(8,8\right)\left(1/2\right)}
	\right|K^{+}\right\rangle_Q -\frac{8}{f^{2}}
	c_{88,4}m_{xz}^{2}\ .
	\label{eq:O88 K->pi sub}
\eea
These relations hold for either $Q=Q_1,Q_2$, and lead to
\begin{eqnarray}
	 \left\langle \pi^{+}\left|\mathcal{O}^{\left(8,8\right)(1/2)}_{\rm sub}
	 \right|K^{+}\right\rangle _{Q_{1}}
	& = & \frac{8\alpha_{88}}{f^{2}}
	+ \frac{4}{f^{2}}\biggl\{
    \left(-c_{88,1}+c_{88,2}+2c_{88,3}\right)m_{xz}m_{X}
    +4c_{88,4}\left(m_{xz}^{2}+m_{X}^{2}\right)
    \nonumber \\	&&{}
    +4c_{88,5}\left(m_{xz}^{2}+m_{X}^{2}\right)   
    +4c_{88,6}\left(2m_D^2+m_S^2\right)
  	\biggr\}
	\nonumber \\
 	&&{} + \left\langle \pi^{+}
	\left|\mathcal{O}^{\left(8,8\right)(1/2)}\right|K^{+}
	\right\rangle _{Q_{1},{\rm logs}}  
    \label{eq:O88 Kpi 1/2 Q1 ct}
\end{eqnarray}
\begin{eqnarray}
	\left\langle \pi^{+}\left|
	\mathcal{O}^{\left(8,8\right)(1/2)}_{\rm sub}
	\right|K^{+}\right\rangle _{Q_{2}}
	& = & \left\langle \pi^{+}\left|
	\mathcal{O}^{\left(8,8\right)(1/2)}_{\rm sub}
	\right|K^{+}\right\rangle _{Q_{1}}
	\nonumber \\&&{}
	+ \frac{4}{3}\frac{\alpha_{88}}{f^{4}}
	\biggl\{
	3\frac{m_{xz}}{m_{xz}-m_{X}}\biggl[\ell(m_{X}^{2}) 
	+\ell(m_{Z}^{2})
	- 2\ell(m_{xz}^{2})\biggr]
	\nonumber \\&&{}
	+ 6m_{xz}m_{X}(\beta(q^{2},m_{xz}^{2},m_{X}^{2})
	-\beta(q^{2},m_{xz}^{2},m_{Z}^{2}))
	\biggr\}\ .
	\label{eq:O88 Kpi 1/2 Q2}
\end{eqnarray}
The logarithms appearing in Eq.~(\ref{eq:O88 Kpi 1/2 Q1 ct}) are given in Appendix~\ref{sub:logterms 88}.

When we have degenerate valence quark masses ($m_{x}=m_{y}=m_{z}$),
the above formula can be simplified. However, for some terms, especially those which involve residue functions $R_{X}\left(m_{a},m_{b}\right)$, taking this limit is non-trivial. Thus, we give the degenerate valence $K\to\pi$ amplitudes explicitly. (The degenerate valence $K\to 0$
matrix elements vanish due to CPS symmetry \cite{Bernard:1989nb}.)
The subtracted amplitude for the degenerate case is given by
\begin{eqnarray}
 	\left\langle \pi^{+}\left|\mathcal{O}^{\left(8,8\right)
	\left(1/2\right)}_{\rm sub}\right|K^{+}
	\right\rangle _{Q}^{\rm deg.val.}
	& = & \left\langle \pi^{+}\left|\mathcal{O}^{\left(8,8\right)
	\left(1/2\right)}\right|K^{+}\right\rangle ^{\rm deg.val.}_Q
	+\frac{4c_{88,4}(2B_0m_x)}
	{\alpha^{\left(3,\bar{3}\right)}}\left\langle \pi^{+}
	\left|\Theta^{\left(3,\bar{3}\right)}\right|K^{+}
	\right\rangle \nonumber \\
	& = & \left\langle \pi^{+}\left|\mathcal{O}^{\left(8,8\right)
	\left(1/2\right)}\right|K^{+}\right\rangle ^{\rm deg.val.}_Q
	-\frac{8}{f^{2}}c_{88,4}m_{X}^{2}\ ,
	\label{eq:O88 K->pi sub del val}
\end{eqnarray}
where again the second equality is correct through NLO in $\chi$PT.  In the degenerate case the amplitudes are the same for $Q_1$ and $Q_2$,
\begin{align}
	\left\langle \pi^{+}\left|\mathcal{O}^{\left(8,8\right)(3/2)}
	\right|K^{+}\right\rangle _{Q_{1,2}}^{\rm deg.val.}= & 
	\frac{4\alpha_{88}}{f^{2}}\left(1+\delta Z_{X}\right)
	\nonumber \\ & +\frac{4\alpha_{88}}{f^{4}}
	\biggl\{-\frac{8}{3}\left[2\ell\left(m_{xd}^{2}\right)
	+\ell\left(m_{xs}^{2}\right)\right]
	+2m_{X}^{2}\tilde{\ell}\left(m_{X}^{2}\right)\biggr\}
	\nonumber \\& +\frac{4}{f^{2}}\bigl[
     \left(-c_{88,1}-c_{88,2}+4c_{88,4}+4c_{88,5}\right)m_{X}^{2}
    \nonumber \\& +2c_{88,6}\left(2m_D^2+m_S^2\right)
     \bigr], \label{eq:O88 kpi deg val 3/2 Q1}\\
\left\langle \pi^{+}\left|\mathcal{O}^{\left(8,8\right)(1/2)}_{sub}\right|K^{+}\right\rangle _{Q_{1,2}}^{\rm deg.val.}= & 
	\frac{8\alpha_{88}}{f^{2}}\left(1+\delta Z_{X}\right)\nonumber \\
 	& +\frac{8\alpha_{88}}{f^{4}}\biggl\{
	-\frac{8}{3}\left[2\ell\left(m_{xd}^{2}\right)
	+\ell\left(m_{xs}^{2}\right)\right]
	-m_{X}^{2}\tilde{\ell}\left(m_{X}^{2}\right)
	\nonumber \\
 	& +\frac{4}{f^{2}}\biggl[
    \left(-c_{88,1}+c_{88,2}+2c_{88,3}
    +8c_{88,4}+8c_{88,5}\right)m_{X}^{2} \nonumber\\
 	&  \qquad
    +4c_{88,6}\left(2m_D^2+m_S^2\right)
    \biggl]. \label{eq:O88 kpi del val 1/2 Q1}  
\end{align}

\subsection{$K\to \pi\pi$ amplitudes in full QCD}

The LEC's needed to construct the (8,8), $K\to\pi\pi$ amplitudes at physical kinematics through NLO can be obtained from the $K\to\pi$ and $K\to 0$ amplitudes given above.  The extraction of LEC's is essentially unchanged from the case of 3 degenerate sea quarks treated in Ref.~\cite{Laiho:2003uy}.  For completeness, we present the physical $K\to\pi\pi$ amplitudes, which were calculated originally in Refs.~\cite{Cirigliano:1999pv,Cirigliano:2001hs}, and subsequently checked in Refs.~\cite{Pallante:2000pz, Laiho:2003uy}. 
\bea \label{eq:K2pipi883/2}
	\langle \pi^+ \pi^- |{\cal O}^{(8,8),(3/2)}| 
	K^0\rangle = -\frac{4i\alpha_{88}}{f_K f_\pi^2}
	+\frac{12i}{f_K f_\pi^2}\left[(-c_{88,2}-c_{88,3}
	-2c_{88,4}-2c_{88,5}-4c_{88,6})m_K^2 \right. \nonumber && \\ 
	\left. -(-c_{88,1}-c_{88,2}+4c_{88,4}+4c_{88,5}+2c_{88,6})
	m_\pi^2\right] + \langle \pi^+ \pi^- |{\cal O}^{(8,8),(3/2)}|
	 K^0\rangle_{\rm logs}, \nonumber \\ &&
\eea
\bea \label{eq:K2pipi881/2}
	\langle \pi^+ \pi^- |{\cal O}^{(8,8),(1/2)}| K^0\rangle 
	= -\frac{8i\alpha_{88}}{f_K f_\pi^2}-\frac{12i}{f_K f_\pi^2}
	\left[(-c_{88,1}-c_{88,2}+4c_{88,4}+4c_{88,5}+8c_{88,6})m_K^2 
	\right. \nonumber && \\ \left. +(-c_{88,1}+c_{88,2}+2c_{88,3}
	+8c_{88,4}+8c_{88,5}+4c_{88,6})m_\pi^2\right] + \langle \pi^+
	 \pi^- |{\cal O}^{(8,8),(1/2)}| K^0\rangle_{\rm logs}. 
	 \nonumber \\ &&
\eea
The logarithmic terms are given in Appendix~\ref{sub:logterms 88}.  Note that in Eqs.~(\ref{eq:K2pipi883/2}) and (\ref{eq:K2pipi881/2}), the decay constants appearing in the tree-level terms are the physical decay constants (correct to one loop).  When constructing $K\to\pi\pi$ amplitudes using Eqs.~(\ref{eq:K2pipi883/2}) and (\ref{eq:K2pipi881/2}), one should use the physical decay constants in the tree-level expression, as determined from lattice calculations or experiment, in order to avoid double counting a subset of the one-loop corrections.

\section{$K\to\pi$ for the $(27,1)$, $\Delta I=3/2$ case}\label{sec:O271wme32}

	The operators which transform as (27,1)'s under the irreducible representation of the chiral symmetry group give the dominant contribution to $\textrm{Re}(A_2)$, \emph{i.e.}, the real part of the $\Delta I=3/2$, $K\to\pi\pi$ amplitude.  In this section we review the (27,1) chiral operators that are needed through NLO, and we give results for the NLO $\Delta I=3/2$, $K\to\pi$ amplitude.

\subsection{Definition of the $\mathcal{O}^{\left(27,1\right)}$
operators\label{sub:O27def}}

Following \cite{Bernard:1985wf,Laiho:2003uy}, the operator in the
(27,1) representation can be written as 
\begin{align}
\mathcal{O}^{\left(27,1\right)}= & T_{kl}^{ij}\left(\Sigma\partial_{\mu}\Sigma^{\dagger}\right)_{i}^{k}\left(\Sigma\partial^{\mu}\Sigma^{\dagger}\right)_{j}^{l}\nonumber \\
 & +c_{27,1}T_{kl}^{ij}\left(S\right)_{i}^{k}\left(S\right)_{j}^{l}\nonumber \\
 & +c_{27,2}T_{kl}^{ij}\left(P\right)_{i}^{k}\left(P\right)_{j}^{l}\nonumber \\
 & +c_{27,4}T_{kl}^{ij}\left(L_{\mu}\right)_{i}^{k}\left(\left\{ L^{\mu},S\right\} \right)_{j}^{l}\nonumber \\
 & +c_{27,5}T_{kl}^{ij}\left(L_{\mu}\right)_{i}^{k}\left(\left[L^{\mu},P\right]\right)_{j}^{l}\nonumber \\
 & +c_{27,6}T_{kl}^{ij}\left(S\right)_{i}^{k}\left(L^{2}\right)_{j}^{l}\nonumber \\
 & +c_{27,7}T_{kl}^{ij}\left(L_{\mu}\right)_{i}^{k}\left(L^{\mu}\right)_{j}^{l}\mathrm{str}\left[S\right]\nonumber \\
 & +c_{27,20}T_{kl}^{ij}\left(L_{\mu}\right)_{i}^{k}\left(\partial_{\nu}W^{\mu\nu}\right)_{j}^{l}\nonumber \\
 & +c_{27,24}T_{kl}^{ij}\left(W_{\mu\nu}\right)_{i}^{k}\left(W^{\mu\nu}\right)_{j}^{l}\label{eq:O27}\end{align}
where $S,P,L_\mu$ are defined in Eq.~(\ref{eq:NLO_pieces}) and 
\begin{equation}
	W_{\mu\nu}=2\left(\partial_{\mu}L_{\nu}
	+\partial_{\nu}L_{\mu}\right).
	\label{eq:W_mu_nu}
\end{equation}
The tensor $T_{kl}^{ij}$ has different elements depending on which isospin we are projecting.  To
project the $\Delta I=3/2$ operator, we set 
\begin{align}
 	& T_{12}^{13}=T_{12}^{31}=T_{21}^{13}=T_{21}^{31}=\frac{1}{2}
	\nonumber \\
 	& T_{22}^{23}=T_{22}^{32}=-\frac{1}{2}\ ,
	\label{eq:O27_DeltaI_32}
\end{align}
whereas for the $\Delta I=1/2$ operator, we set 
\begin{align}
	 & T_{12}^{13}=T_{12}^{31}=T_{21}^{13}=T_{21}^{31}=\frac{1}{2}
	 \nonumber \\
 	& T_{22}^{23}=T_{22}^{32}=1\nonumber \\
 	& T_{32}^{33}=T_{23}^{33}=-\frac{3}{2}\ .
	\label{eq:O27_DeltaI_12}
\end{align}

In order to adapt Eq.~(\ref{eq:O27}) to the partially quenched theory, we must promote $T$ to a $9^4$ element tensor, although many components will remain zero (only the $3^4$ block corresponding to the valence quark sector will have non-zero elements). To take into account the graded nature of the group, we multiply by factors of $\varepsilon_i$ defined in Eq.~(\ref{eq:propsign}),
such that
\begin{equation}
	\mathcal{O}^{\left(27,1\right)}
	=\sum_{ijkl}\varepsilon_i\varepsilon_j
	T_{kl}^{ij}
	\left(\Sigma\partial_{\mu}\Sigma^{\dagger}\right)_{i}^{k}
	\left(\Sigma\partial^{\mu}\Sigma^{\dagger}\right)_{j}^{l}\ ,
	\label{eq:O27-super}
\end{equation}
where we display the summation over $i,j,k,l$ explicitly for clarity. 

There is another equivalent approach to obtaining the partially quenched operators for the (27,1) case.
It is possible, as illustrated in Appendix D of \cite{Blum:2001xb}, to rewrite Eq.~(\ref{eq:O27})  in terms of traces over the various operators.
The partially quenched theory is then obtained in the usual way by changing traces to
supertraces, and we obtain
\begin{eqnarray}
  O^{(27,1),(3/2)} &=& 
  \Str[\lambda_6\Sigma \partial_\mu \Sigma^\dagger]
  \Str[A\Sigma \partial^\mu \Sigma^\dagger]
  +
  \Str[\lambda_3\Sigma \partial_\mu \Sigma^\dagger]
  \Str[\lambda_4\Sigma \partial^\mu \Sigma^\dagger],\\
  O^{(27,1),(1/2)} &=& 
  \Str[\lambda_6\Sigma \partial_\mu \Sigma^\dagger]
  \Str[B\Sigma \partial^\mu \Sigma^\dagger]
  +
  \Str[\lambda_3\Sigma \partial_\mu \Sigma^\dagger]
  \Str[\lambda_4\Sigma \partial^\mu \Sigma^\dagger]\ ,
\end{eqnarray}
where we have defined the matrices
\begin{eqnarray}
  (\lambda_3)_{ij} = \delta_{i3}\delta_{j1},&\quad&
  (\lambda_4)_{ij} = \delta_{i1}\delta_{j2},\\
  A_{ij} = \delta_{i1}\delta_{j1} - \delta_{i2}\delta_{j2},&\quad&
  B = {\rm diag}(1,2, -3).
\end{eqnarray}

Since the kaon has isospin $I=1/2$, the $\Delta I=3/2, K\to0$ process vanishes. The amplitude for the (27,1), $\Delta I = 3/2$, $K\to\pi$ matrix element is
\begin{eqnarray}
 	\left\langle \pi^{+}\left|\mathcal{O}^{\left(27,1\right)(3/2)}
	\right|K^{+}\right\rangle 
	& = & -\frac{4\alpha_{27}}{f^{2}}m_{X}m_{xz}
	+ \frac{1}{f^{2}}\biggl[
    16\left(-c_{27,2}+4c_{27,24}\right)m_{X}^{2}m_{xz}^{2}
    \nonumber \\&&{}
    +8\left(c_{27,4}-c_{27,20}\right)m_{X}m_{xz}
    \left(m_{X}^{2}+m_{xz}^{2}\right)
    \nonumber \\&&{}
    +8c_{27,7}m_{X}m_{xz}\left(2m_D^2+m_S^2\right)
    \biggr]   
	+ \left\langle \pi^{+}\left|\mathcal{O}^{\left(27,1\right)(3/2)}
	\right|K^{+}\right\rangle_{\rm logs} ,                                                   \label{eq:O27 Kpi 3/2 ct}
\end{eqnarray}
where the logarithmic terms are given in Appendix~\ref{sub:logterms 271}.

For degenerate valence quarks, this amplitude simplifies to
\begin{eqnarray}
	\left\langle \pi^{+}\left|\mathcal{O}^{\left(27,1\right)(3/2)}
	\right|K^{+}\right\rangle ^{\rm deg.val.}
	& = & -\frac{4\alpha_{27}}{f^{2}}m_{X}^{2}\left[1+
	\delta Z_{X}+\frac{\delta m_{X}^{2}}{m_{X}^{2}}\right]
	\nonumber \\&&{}
	+ \frac{8}{3}\frac{\alpha_{27}}{f^{4}}m_{X}^{2}\biggl\{
	6\ell(m_{X}^{2})+8\ell(m_{xd}^{2})+4\ell(m_{xs}^{2})
	-3m_{X}^{2}\tilde{\ell}(m_{X}^{2})\biggr\}
	\nonumber \\&&{}
	+\frac{1}{f^{2}}\biggl[
    16\left(-c_{27,2}+c_{27,4}-c_{27,20}+4c_{27,24}\right)m_{X}^{4}
	\nonumber \\&&{}
    +8c_{27,7}m_{X}^{2}\left(2m_D^2+m_S^2\right)
    \biggr]\ .
\label{eq:O27 Kpi degval 3/2}
\end{eqnarray}

\section{$\Delta I=1/2$ Weak Matrix Elements 
for (8,1) and (8,1)+(27,1) operators}\label{sec:O81_271_12}

In this section we present results for $\Delta I=1/2$ amplitudes, which include $K\to 0$ and $K\to~\pi$ matrix elements, for operators that transform under the $(8,1)+(27,1)$ representation, and for those that transform under the pure (8,1) representation.  We perform the power subtraction explicitly through NLO in the chiral expansion and present the subtracted amplitudes.

\subsection{Definition of the $\mathcal{O}^{\left(8,1\right)}$ operator}

As shown in Ref. \cite{Bernard:1985wf}, there are two leading order [$\mathcal{O}\left(p^{2}\right)$]
operators in the chiral symmetry group (8,1), with coefficients $\alpha_{1}$ and $\alpha_{2}$.  There
are 13 NLO [$\mathcal{O}\left(p^{4}\right)$] operators relevant for this work.  Note that there is an extra operator which only appears in the partially quenched case.  In the full theory, by convention \cite{Kambor:1989tz}, operator 14 is absorbed into operators 10, 11, 12, and 13 via the Cayley-Hamilton theorem. This is not possible in the partially quenched theory \cite{Laiho:2003uy, Sharpe:2003vy}. 
The (8,1) operators are
\begin{align}
\mathcal{O}^{\left(8,1\right)}= &
	\alpha_{1}\mathrm{str}\left[\lambda_{6}\partial_{\mu}\Sigma\partial^{\mu}\Sigma^{\dagger}\right]
	+\alpha_{2}2B_{0}\mathrm{str}\left[\lambda_{6}
		\left(\cM^{\dagger}\Sigma^{\dagger}+\Sigma \cM \right) \right]     \nonumber \\
 	& +c_{81,1}\mathrm{str}\left[\lambda_{6}S^{2}\right]
	+c_{81,2}\mathrm{str}\left[\lambda_{6}S\right]\mathrm{str}\left[S\right]\nonumber \\
 	& +c_{81,3}\mathrm{str}\left[\lambda_{6}P^{2}\right]
	+c_{81,4}\mathrm{str}\left[\lambda_{6}P\right]\mathrm{str}\left[P\right]\nonumber \\
 	& +c_{81,5}\mathrm{str}\left[\lambda_{6}\left[ P, S \right]\right]
	+c_{81,10}\mathrm{str}\left[\lambda_{6}\left\{ S,L^{2}\right\} \right]\nonumber \\
 	& +c_{81,11}\mathrm{str}\left[\lambda_{6}L_{\mu}SL^{\mu}\right]
	+c_{81,12}\mathrm{str}\left[\lambda_{6}L_{\mu}\right]
		\mathrm{str}\left[\left\{ L^{\mu},S\right\} \right]\nonumber \\
 	& +c_{81,13}\mathrm{str}\left[\lambda_{6}S\right]\mathrm{str}\left[L^{2}\right]
	+c_{81,14}\mathrm{str}\left[\lambda_{6}L^{2}\right]\mathrm{str}\left[S\right]\nonumber \\
 	& +c_{81,15}\mathrm{str}\left[\lambda_{6}\left[ L^{2}, P \right]\right]
	+c_{81,35}\mathrm{str}\left[\lambda_{6}
		\left\{ L_{\mu},\partial_{\nu}W^{\mu\nu}\right\} \right]\nonumber \\
 	& +c_{81,39}\mathrm{str}\left[\lambda_{6}W_{\mu\nu}W^{\mu\nu}\right]      \label{eq:O81}
\end{align}
The (27,1) operators relevant for this section are given in Eq.~(\ref{eq:O27}).

\subsection{$K\to 0$ amplitudes}

The NLO expression for the unsubtracted $K\to 0$ amplitude in the pure (8,1) case is 
\begin{eqnarray}
	 \left\langle 0\left|\mathcal{O}^{\left(8,1\right)}\right|K^{0}
	 \right\rangle 
	& = & \frac{4i}{f}\alpha_{2}(m_{xz}^{2}-m_{X}^{2})
	+ \frac{8i}{f}\left(m_{xz}^{2}-m_{X}^{2}\right)\bigl[
    2\left(c_{81,1}-c_{81,5}\right)m_{xz}^{2}
	\nonumber \\&&{}
 	+c_{81,2}\left(2m_D^2+m_S^2\right)
	\bigr]
	+ \left\langle 0\left|\mathcal{O}^{\left(8,1\right)}
	\right|K^{0}\right\rangle_{\rm logs} ,    
   \label{eq:O81 Kvac ct}
\end{eqnarray}
where the logarithmic terms are given in Appendix \ref{sub:logterms 81}.  For the (8,1)+(27,1) case, we have
\begin{eqnarray}
 	\left\langle 0\left|\mathcal{O}^{(8,1)+\left(27,1\right)(1/2)}
	\right|K^{0}\right\rangle
	& = & \left\langle 0\left|\mathcal{O}^{\left(27,1\right)}
	\right|K^{0}\right\rangle_{\rm logs}
		\nonumber \\&&{}
 	+\frac{48i}{f}c_{27,1}\left(m_{xz}^{2}-m_{X}^{2}\right)^{2} 
	+  \left\langle 0\left|\mathcal{O}^{\left(8,1\right)}\right|K^{0}
	 \right\rangle,         \label{eq:O27 Kvac}
 \end{eqnarray}
where
 \begin{eqnarray}
 	\left\langle 0\left|\mathcal{O}^{\left(27,1\right)}
	\right|K^{0}\right\rangle_{\rm logs}
	& = & 
	\frac{4i}{f^{3}}\alpha_{27}\biggl\{\biggl[2m_{X}^{2}
	-R_{X}(m_{\eta})+m_{X}^{2}R_{\eta}(m_{X},m_{X})
	\nonumber \\&&{}
	+2m_{X}^{2}R_{X}(m_{\eta},m_{Z})\biggr]\ell(m_{X}^{2})
 	+\biggl[2m_{Z}^{2}-R_{Z}(m_{\eta})
	\nonumber \\&&{}
	+m_{Z}^{2}R_{\eta}(m_{Z},m_{Z})
	+2m_{Z}^{2}
	R_{Z}(m_{\eta},m_{X})\biggr]\ell(m_{Z}^{2})
	\nonumber \\&&{}
 	-6m_{xz}^{2}\ell(m_{xz}^{2})
 	+\biggl[-m_{\eta}^{2}R_{\eta}(m_{Z},m_{Z})
	+2m_{\eta}^{2}R_{\eta}(m_{X},m_{Z})
	\nonumber \\&&{}
	-m_{\eta}^{2}
	R_{\eta}(m_{X},m_{X})\biggr]\ell(m_{\eta}^{2})
 	+m_{X}^{2}R_{X}(m_{\eta})\tilde{\ell}(m_{X}^{2})
	\nonumber \\&&{}
	+m_{Z}^{2}
	R_{Z}(m_{\eta})\tilde{\ell}(m_{Z}^{2})\biggr\}.
	\nonumber \\&&{}
          \label{eq:O27 Kvaclogs}
 \end{eqnarray}

Following the procedure given in Ref.~\cite{Blum:2001xb},
we perform the subtraction of  the power divergence in (8,1) and (8,1)+(27,1) amplitudes.  In order to do this we require the amplitudes for $\Theta^{\left(3,\bar{3}\right)}$ through NLO, given in Section \ref{sec:subtractions}.
The ratio of (8,1) and $(3,\overline{3})$, $K\to 0$ matrix elements to NLO is
\begin{align}
\frac{\left\langle 0\left|\mathcal{O}^{\left(8,1\right)}\right|K^{0}\right\rangle }{\left\langle 0\left|\Theta^{\left(3,\bar{3}\right)}\right|K^{0}\right\rangle }= & 2\frac{\alpha_{2}}{\alpha^{\left(3,\bar{3}\right)}}B_0\left(m_{z}-m_{x}\right)+\frac{f}{2i\alpha^{\left(3,\bar{3}\right)}}\langle0|{\cal O}^{(8,1)}|K^0\rangle^{(1)}_{\rm logs}\nonumber \\
 & +\frac{4}{\alpha^{\left(3,\bar{3}\right)}}\left(m_{xz}^{2}-m_{X}^{2}\right)\left[2\left(c'_{81,1}-c'_{81,5}\right)m_{xz}^{2}+c'_{81,2}
 	\left(2m_D^2 + m_S^2\right)
	\right],\label{eq:O81 K->0 ratio1}
\end{align}
where the transformed coefficients $c'_{81,1}$, $c'_{81,2}$, $c'_{81,5}$ [defined in Table~\ref{tab:LECtrans}] are linear combinations of the original LEC's, $c_{81,1}$, etc., and the Gasser-Leutwyler coefficients originating from $\mathcal{O}(p^2)$ terms in the amplitude $\left\langle 0\left|\Theta^{\left(3,\bar{3}\right)}\right|K^{0}\right\rangle$.  The first term on the right-hand side of Eq.~(\ref{eq:O81 K->0 ratio1}) contains the power divergence, which is proportional to $m_z-m_x$ to all orders in the chiral expansion.  The remaining terms are finite, including the rotated LEC's $c'_{81,i}$.  Since the rotated LEC's contain a term proportional to $\alpha_2$, it follows that the unrotated $c_{81,i}$'s must also contain power divergences \cite{Golterman:2004mf}.  This was implicitly assumed in the work of Ref.~\cite{Laiho:2003uy}.

A similar expression exists for the ratio involving the (8,1)+(27,1) amplitude,
\begin{align}
\frac{\left\langle 0\left|\mathcal{O}^{\left(8,1\right)+\left(27,1\right)(1/2)}\right|K^{0}\right\rangle }{\left\langle 0\left|\Theta^{\left(3,\bar{3}\right)}\right|K^{0}\right\rangle }= & 2\frac{\alpha_{2}}{\alpha^{\left(3,\bar{3}\right)}}B_0\left(m_{z}-m_{x}\right)+\frac{f}{2i\alpha^{\left(3,\bar{3}\right)}}\langle0|{\cal O}^{(8,1)}|K^0\rangle^{(1)}_{\rm logs} \nonumber \\ &+\frac{f}{2i\alpha^{\left(3,\bar{3}\right)}}\langle0|{\cal O}^{(27,1)}|K^0\rangle_{\rm logs}\nonumber \\
 & +\frac{4}{\alpha^{\left(3,\bar{3}\right)}}\left(m_{xz}^{2}-m_{X}^{2}\right)\left[2\left(c'_{81,1}-c'_{81,5}\right)m_{xz}^{2}+c'_{81,2}
 	\left(2m_D^2 + m_S^2\right)
	\right. \nonumber  \\  & \left. + 6c_{27,1}(m_{xz}^2-m_X^2) \right].\label{eq:O81 K->0 ratio2}
\end{align}

\begin{table}[tbp]
\caption{The transformation of the (8,1) LEC's (denoted in the text with a prime) under the vacuum subtraction process.} \label{tab:LECtrans}
 \begin{tabular}{c|c}
 \hline
 \hline
 LEC         & Transformed LEC \\
 \hline
 $c_{81,1}$  & $c_{81,1}-(4\alpha_2/f^2)(2L_8+H_2)$ \\
 $c_{81,2}$  & $c_{81,2}-(16\alpha_2/f^2)L_6$ \\
 $c_{81,3}$  & $c_{81,3}+(4\alpha_2/f^2)(-2L_8+H_2)$ \\
 $c_{81,5}$  & $c_{81,5}-(4\alpha_2/f^2)H_2$ \\
 $c_{81,10}$ & $c_{81,10}-(4\alpha_2/f^2)L_5$ \\
 $c_{81,13}$ & $c_{81,13}-(8\alpha_2/f^2)L_4$ \\
 $c_{81,15}$ & $c_{81,15}+(4\alpha_2/f^2)L_5$ \\
 \hline
 \hline
 \end{tabular} 
\end{table}

\subsection{$K\to\pi$ amplitudes with 2+1 valence flavors}

For the $K\to\pi$ amplitudes, we first present the matrix elements of the unsubtracted operators,
\begin{eqnarray}
 	\left\langle \pi^{+}\left|\mathcal{O}^{\left(8,1\right)}
	\right|K^{+}\right\rangle &= & 
	\frac{4\alpha_{1}}{f^{2}}m_{xz}m_{X} 
	- \frac{4\alpha_{2}}{f^{2}}m_{xz}^{2} \nonumber \\
 	&&{}+ \frac{8}{f^{2}}\biggl[
    -2c_{81,1}m_{xz}^{4}
	-c_{81,2}m_{xz}^{2}\left(2m_D^2+m_S^2\right)
	-2c_{81,3}m_{X}^{2}m_{xz}^{2}
	\nonumber \\&&{}
	+2c_{81,5}m_{xz}^{2}\left(m_{xz}^{2}-m_{X}^{2}\right)      	
	+2c_{81,10}m_{X}m_{xz}^{3}
	+c_{81,11}m_{X}^{3}m_{xz}
	\nonumber \\&&{}
	+c_{81,14}m_{X}m_{xz}\left(2m_D^2+m_S^2\right)   
	-2c_{81,35}m_{X}m_{xz}\left(m_{X}^{2}+m_{xz}^{2}\right)
	\nonumber \\&& {}
	+8c_{81,39}m_{X}^{2}m_{xz}^{2}
	\biggr]
	+ \left\langle \pi^{+}\left|\mathcal{O}^{\left(8,1\right)}
	\right|K^{+}\right\rangle_{\rm logs} ,                                           \label{eq:O81 Kpi ct}
\end{eqnarray}
\begin{eqnarray}
 	\left\langle \pi^{+}\left|\mathcal{O}^{(8,1)+\left(27,1\right)(1/2)}
	\right|K^{+}\right\rangle 
	&= & 
	-\frac{4\alpha_{27}}{f^{2}}m_{xz}m_{X}
	+ \frac{1}{f^{2}}\biggl[
    48c_{27,1}m_{xz}^{2}\left(m_{xz}^{2}-m_{X}^{2}\right)
    \nonumber \\&&{}
    +16\left(-c_{27,2}+4c_{27,24}\right)m_{X}^{2}m_{xz}^{2}  
    \nonumber \\ && +8\left(c_{27,4}-c_{27,20}\right)m_{X}m_{xz}\left(m_{X}^{2}
	+m_{xz}^{2}\right)
    \nonumber \\&&{}
    +24c_{27,6}m_{X}m_{xz}\left(m_{X}^{2}-m_{xz}^{2}\right)
    +8c_{27,7}m_{X}m_{xz}\left(2m_D^2+m_S^2\right)
    \biggr]          
    \nonumber \\&&{}
    + \left\langle \pi^{+}\left|
    \mathcal{O}^{\left(27,1\right)(1/2)}\right|K^{+}
    \right\rangle_{\rm logs} + \left\langle \pi^{+}\left|\mathcal{O}^{\left(8,1\right)}
	\right|K^{+}\right\rangle,
    \label{eq:O27 Kpi 1/2 ct}
\end{eqnarray}
where the logarithmic contributions to the (8,1) amplitude are given in Appendix \ref{sub:logterms 81}, and the logarithmic contributions to the (27,1) amplitude are given in Appendix \ref{sub:logterms 271}.

Given the coefficient of the power divergent term from $K\to 0$, it is possible to carry out the operator subtraction in $K\to \pi$ numerically.  By CPS symmetry, the following subtraction removes the divergence to all orders in the chiral expansion,
\begin{eqnarray}
	 \left\langle \pi^{+}\left|\mathcal{O}_{\rm sub}^{\left(8,1\right)}
	 \right|K^{+}\right\rangle 
	& = & \left\langle \pi^{+}\left|\mathcal{O}^{\left(8,1\right)}
	\right|K^{+}\right\rangle -2\frac{\alpha_{2}}
	{\alpha^{\left(3,\bar{3}\right)}}B_0(m_z+m_x)\left\langle \pi^{+}\left|
	\Theta^{\left(3,\bar{3}\right)}\right|K^{+}\right\rangle.
	\nonumber \\ 
\label{eq:O81 Kpi sub def}
\end{eqnarray}
To NLO this expression becomes
\begin{eqnarray}
	 \left\langle \pi^{+}\left|\mathcal{O}_{\rm sub}^{\left(8,1\right)}
	 \right|K^{+}\right\rangle 
	& = & \frac{4\alpha_{1}}{f^{2}}m_{xz}m_{X}  + \frac{8}{f^{2}}\biggl[
    -2c'_{81,1}m_{xz}^{4}
	-c'_{81,2}m_{xz}^{2}\left(2m_D^2+m_S^2\right) \nonumber \\ &&
	-2c'_{81,3}m_{X}^{2}m_{xz}^{2}
	+2c'_{81,5}m_{xz}^{2}\left(m_{xz}^{2}-m_{X}^{2}\right)      	
	+2c'_{81,10}m_{X}m_{xz}^{3} \nonumber \\&&{}
	+c_{81,11}m_{X}^{3}m_{xz}
	+c_{81,14}m_{X}m_{xz}\left(2m_D^2+m_S^2\right)   \nonumber \\&&{}
	-2c_{81,35}m_{X}m_{xz}\left(m_{X}^{2}+m_{xz}^{2}\right)
	+8c_{81,39}m_{X}^{2}m_{xz}^{2}
	\biggr] \nonumber \\&& {}
+
	\left\langle \pi^{+}\left|\mathcal{O}^{\left(8,1\right)}
	\right|K^{+}\right\rangle^{(1)}_{\rm logs} ,
\label{eq:O81 Kpi sub}
\end{eqnarray}
where again, the $c'_{81,i}$ are the linear combinations of LEC's given in Table~\ref{tab:LECtrans}.  Note that the subtraction eliminates the term in Eq.~(\ref{eq:O81 Kpi ct}) proportional to $\alpha_2$.  The NLO chiral logarithms proportional to $\alpha_2$ are also eliminated.  The remaining logarithms are contained in the term $\left\langle \pi^{+}\left|\mathcal{O}^{\left(8,1\right)}\right|K^{+}\right\rangle^{(1)}_{logs}$ given in Appendix~\ref{sub:logterms 81}; this term is proportional to $\alpha_1$.

A similar subtraction can be performed for the (8,1)+(27,1) case,
\begin{eqnarray}
	 \left\langle \pi^{+}\left|\mathcal{O}_{\rm sub}^{\left(8,1\right)+(27,1)(1/2)}
	 \right|K^{+}\right\rangle 
	& = & \left\langle \pi^{+}\left|\mathcal{O}^{\left(8,1\right)+(27,1)(1/2)}
	\right|K^{+}\right\rangle \nonumber \\ && -2\frac{\alpha_{2}}
	{\alpha^{\left(3,\bar{3}\right)}}B_0(m_z+m_x)\left\langle \pi^{+}\left|
	\Theta^{\left(3,\bar{3}\right)}\right|K^{+}\right\rangle.
	\nonumber \\ 
\label{eq:O81
+27 Kpi sub def}
\end{eqnarray}
Again, this subtraction removes the power divergences to all orders in the chiral expansion.  To NLO the subtracted operator gives the matrix element,
\begin{eqnarray}
 	\left\langle \pi^{+}\left|\mathcal{O}_{\rm sub}^{(8,1)+\left(27,1\right)(1/2)}
	\right|K^{+}\right\rangle 
	&= & 
	-\frac{4\alpha_{27}}{f^{2}}m_{xz}m_{X}
	+ \frac{1}{f^{2}}\biggl[
    48c_{27,1}m_{xz}^{2}\left(m_{xz}^{2}-m_{X}^{2}\right)
    \nonumber \\&&{}
    +16\left(-c_{27,2}+4c_{27,24}\right)m_{X}^{2}m_{xz}^{2}  
    \nonumber \\ && +8\left(c_{27,4}-c_{27,20}\right)m_{X}m_{xz}\left(m_{X}^{2}
	+m_{xz}^{2}\right)
    \nonumber \\&&{}
    +24c_{27,6}m_{X}m_{xz}\left(m_{X}^{2}-m_{xz}^{2}\right)
    +8c_{27,7}m_{X}m_{xz}\left(2m_D^2+m_S^2\right)
    \biggr]          
    \nonumber \\&&{}
    + \left\langle \pi^{+}\left|
    \mathcal{O}^{\left(27,1\right)(1/2)}\right|K^{+}
    \right\rangle_{\rm logs} + \left\langle \pi^{+}\left|\mathcal{O}_{\rm sub}^{\left(8,1\right)}
	\right|K^{+}\right\rangle.
    \label{eq:O81+27 Kpi 1/2 sub}
\end{eqnarray}

For degenerate valence quarks, Eq.~(\ref{eq:O81 Kpi sub}) reduces to
\begin{eqnarray}
 	\left\langle \pi^{+}\left|\mathcal{O}_{\rm sub}^{\left(8,1\right)}
	\right|K^{+}\right\rangle ^{\rm deg.val.}
	& = & \frac{4}{f^{2}}m_{X}^{2}\Biggl\{\alpha_{1}
	+ 2\biggl[
    \bigl(-2c'_{81,1}-2c'_{81,3}+2c'_{81,10}
	+c_{81,11}
    	\nonumber \\&&{}
    -4c_{81,35}+8c_{81,39}\bigr)m_{X}^2    
     +\left(-c'_{81,2}+c_{81,14}\right)
     \left(2m_D^2+m_S^2\right)
	\biggr]\Biggr\}
	\nonumber \\&&{}
	+ \left\langle \pi^{+}\left|\mathcal{O}^{\left(8,1\right)}
	\right|K^{+}\right\rangle ^{\rm deg.val.,(1)}_{logs},                                                           \label{eq:O81 Kpi deg val ct}
\end{eqnarray}
where the logarithmic terms are given in Appendix \ref{sub:logterms 81}. For
the $(8,1)+(27,1)$, $K\to\pi$ matrix element, we have for the degenerate valence case,
\begin{eqnarray}
	\left\langle \pi^{+}\left|\mathcal{O}_{\rm sub}^{(8,1)+\left(27,1\right)(1/2)}
	\right|K^{+}\right\rangle ^{\rm deg.val.}
	& = & -\frac{4\alpha_{27}}{f^{2}}m_{X}^{2}\left[1+
	\delta Z_{X}+\frac{\delta m_{X}^{2}}{m_{X}^{2}}\right]
	\nonumber \\&&{}
	+ \frac{8}{3}\frac{\alpha_{27}}{f^{4}}m_{X}^{2}\biggl\{
	6\ell(m_{X}^{2})+8\ell(m_{xd}^{2})
	\nonumber \\&&{}
	+4\ell(m_{xs}^{2})
	-3m_{X}^{2}\tilde{\ell}(m_{X}^{2})\biggr\}
	\nonumber \\&&{}
	+\frac{1}{f^{2}}\biggl[
    16\left(-c_{27,2}+c_{27,4}-c_{27,20}+4c_{27,24}\right)m_{X}^{4}
	\nonumber \\&&{}
    +8c_{27,7}m_{X}^{2}\left(2m_D^2+m_S^2\right)
    \biggr] 
   	\nonumber \\&&{}
	+ \left\langle \pi^{+}\left|\mathcal{O}_{sub}^{\left(8,1\right)}
	\right|K^{+}\right\rangle ^{\rm deg.val.} .
\label{eq:O27 Kpi degval 1/2}
\end{eqnarray}
Note that for degenerate valence quarks the (27,1), $\Delta I=1/2$ amplitude is the same as the $(27,1),\Delta I=3/2$ amplitude, Eq.~(\ref{eq:O27 Kpi degval 3/2}).

\section{Finite Volume Corrections}\label{sec:FV}
 
Incorporating the leading corrections coming from the finite volume used in lattice simulations for the above expressions is straightforward.  Here we assume that the time extent used to extract the above matrix elements is infinite, and that the only corrections come from the finite spatial volume. There are two classes of one-loop integrals that must be replaced by their finite volume counterparts. The first is defined in Eq.~(\ref{eq:LoopFunc}), and its associated double-pole counterparts are defined in Eqs.~(\ref{eq:LoopTilde}) and (\ref{eq:LoopTildeTilde}) [these are related to Eq.~(\ref{eq:LoopFunc}) by derivatives with respect to $m^2$]. As discussed in Refs.~\cite{Bernard:2001yj,Aubin:2003mg}, finite volume effects can be accounted for by making the replacements
\begin{equation}\label{eq:ellFV}
	\ell(m^2) \to \ell(m^2) + \frac{1}{16\pi^2}m^2\delta_1(mL)\ ,
\end{equation}
\begin{equation}\label{eq:elltildeFV}
	 \tilde \ell(m^2) \to  \tilde\ell(m^2) + \frac{1}{16\pi^2}
	 \delta_3(mL) \ ,
\end{equation}
\begin{equation}\label{eq:elltildetildeFV}
	 \tilde{\tilde{\ell}}(m^2)
	  \to  \tilde{\tilde{\ell}}(m^2)
	  + \frac{1}{16\pi^2}\frac{\delta_5(mL)}{m^2} \ ,
\end{equation}
with
\begin{equation}\label{eq:delta1}
	\delta_1(mL)  =  4
		\sum_{\mathbf{n}\ne 0}
		\frac{K_1(|\mathbf{n}|mL)}{|\mathbf{n}|mL} \ ,
\end{equation}
\begin{equation}\label{eq:delta3}
	\delta_3(mL)  = -\frac{\partial}{\partial m^2}
	\left[m^2\delta_1(mL)\right]
	=
	 2 \sum_{\mathbf{n}\ne 0}K_0(|\mathbf{n}|mL)\ ,
\end{equation}
\begin{equation}\label{eq:delta5}
	\delta_5(mL)  = m^2\frac{\partial}{\partial m^2}
	\left[\delta_3(mL)\right]
	=
	 - \sum_{\mathbf{n}\ne 0}
	(|\mathbf{n}|mL)
	 K_1(|\mathbf{n}|mL)\ ,
\end{equation}
with $K_0,K_1$ the modified Bessel functions of imaginary argument. 

The second class of loop integrals are more complicated, and are defined in Eqs.~(\ref{eq:BLoopFunc}) and (\ref{eq:beta_tilde_LoopFunc}). For these, we recall the technique used to calculate the above finite volume corrections.  We begin with the finite volume Euclidean space version of Eq.~(\ref{eq:BLoopFunc}), and apply the Poisson Resummation Formula (as discussed in Refs.~\cite{Bernard:2001yj,Arndt:2004bg}).  This leads to the following replacements,
\begin{equation}
	\beta\left(q^{2},m_{1}^{2},m_{2}^{2}\right) 
	\to
	\beta\left(q^{2},m_{1}^{2},m_{2}^{2}\right) 
	+\frac{1}{4\pi^2}\delta_{\beta}(qL, m_1L, m_2L)
	\label{eq:betaFV}\ ,
\end{equation}
\begin{eqnarray}
	\tilde{\beta}\left(q^{2},m_{1}^{2},m_{2}^{2}\right)
	& \to &
	\tilde{\beta}\left(q^{2},m_{1}^{2},m_{2}^{2}\right)
	-
	\frac{1}{4\pi^2 m_1^2}\delta_{\tilde\beta}(qL, m_1L, m_2L),
	\label{eq:betatildeFV}
\end{eqnarray}
and the corrections
\begin{equation}
	\delta_{\beta}(qL, m_1L, m_2L)
	\equiv
	\sum_{\mathbf{n}\ne0}
	\int_0^\infty d k
	\frac{k\sin(k|\mathbf{n}|)}{|\mathbf{n}|}
	\frac{\omega_1+\omega_2}
	{\omega_1\omega_2[(qL)^2 + (\omega_1+\omega_2)^2]}\ ,
	\label{eq:FVbeta_LoopFunc}
\end{equation}
\begin{equation}
	\delta_{\tilde\beta}(qL, m_1L, m_2L)
	\equiv
	(m_1L)^2
	\sum_{\mathbf{n}\ne0}
	\int_0^\infty d k
	\frac{k \sin(k|\mathbf{n}|)}{|\mathbf{n}|}
	\frac{
	(qL)^2\omega_2
	+ (2\omega_1+\omega_2)(\omega_1+\omega_2)^2
	}
	{2\omega_1^3\omega_2
	[(qL)^2 + (\omega_1+\omega_2)^2]^2}
	\ ,
	\label{eq:FVbeta_LoopFunc_2}
\end{equation}
where we have defined
\[
	\omega_i = \sqrt{k^2 + (m_iL)^2}\ ,
\]
and where the function in Eq.~(\ref{eq:FVbeta_LoopFunc_2}) is obtained by taking the partial derivative of Eq.~(\ref{eq:FVbeta_LoopFunc}) with respect to $m_1^2$.

These formulas can be simplified as in Ref.~\cite{Arndt:2004bg}, but only in special cases (such as degenerate masses). For the general case, it is more difficult to find an approximate expression for these finite volume corrections.\footnote{One cannot apply the expansion in Ref.~\cite{Arndt:2004bg}, for example, because these integrals have three different relevant scales, as given by $q^2$, $m_1^2$, and $m_2^2$.} However, it is relatively simple to evaluate these expressions numerically at a finite number of points. Given a set of lattice data at a number of quark masses and lattice volumes, it would be straightforward to tabulate the appropriate finite volume corrections from the above formulas.

\section{Conclusions}\label{sec:conc}

This paper presents the calculation of $K\to 0$ and $K\to\pi$ amplitudes to NLO in PQ$\chi$PT with 2+1 flavors of non-degenerate sea quarks.  Results are presented for both the $\Delta I=1/2$ and $3/2$ channels, for chiral operators corresponding to current-current, gluonic penguin, and electroweak penguin 4-quark operators.  The chiral operators are conveniently grouped by their chiral transformation properties; this work computes matrix elements of (8,8), (27,1), (8,1), and (8,1)+(27,1) chiral operators.  The power divergent operator subtraction is performed explicitly through NLO in the chiral expansion for $\Delta I=1/2$ matrix elements.  We have also shown how to include finite volume effects through one-loop for the quantities considered in this work.   These results are useful for studying the chiral behavior of currently available 2+1 flavor lattice QCD results \cite{Li:2007bx}, from which the low energy constants of the chiral effective theory can be determined.  The low energy constants of these matrix elements are necessary for an understanding of the $\Delta I=1/2$ rule and for calculations of $\epsilon'/\epsilon$ using current lattice QCD simulations.  Electroweak penguin $K\to\pi\pi$ matrix elements can be constructed to NLO in $\chi$PT using the formulas presented in this work, allowing the convergence of the chiral expansion to be studied.  This will serve as a useful cross-check for other non-$\chi$PT methods such as those proposed in Refs.~\cite{Lellouch:2000pv, Buchler:2001np}.

\section*{Acknowledgments}

We thank Norman Christ and Amarjit Soni for discussions, and Ruth Van de Water for comments on the manuscript.
JL acknowledges the hospitality of Columbia University, where part of this work was completed.
This research was supported in part by the U.S.
 DOE under Grant Nos. DE-FG02-92ER40699 (CA and SL), DE-FG02-04ER41302 (CA), DE-AC02-76CH03000 and DE-FG02-91ER40628 (JL), DE-FG02-05ER25681 (MFL), and in part by the NSF under Grant No. PHY-0555235 (JL).  

\appendix

\section{Isospin Decomposition\label{sec:Isospin-Decomposition}}\label{sec:isospindecomp}

The operator that governs the transition $K\to\pi$
can have either isospin 1/2 or isospin 3/2. In typical
lattice calculations, these two processes are calculated independently \cite{Blum:2001xb, Noaki:2001un},
since the operator with isospin 1/2 mixes with a divergent lower dimensional
operator, which must be subtracted.  The isospin 3/2 amplitude
does not have this complication. Therefore, we calculate
the amplitudes for these two processes separately, making use of
the amplitudes for $K^+\to \pi^+$ and $K^{0}\to\pi^{0}$.

We define $\mathcal{M}_{+}=\left\langle \pi^{+}\left|
\mathcal{O}_{i}\right|K^{+}\right\rangle $,
where $\mathcal{O}_{i}$ represents some $\Delta S=1$ operator with
both isospin 1/2 and 3/2 components, and $\mathcal{M}_{0}=\left\langle \pi^{0}\left|\mathcal{O}_{i}\right|K^{0}\right\rangle $.
If we decompose the operator $\mathcal{O}_{i}$ by isospin, $\mathcal{O}_{i}=\mathcal{O}_{i}^{(3/2)}+\mathcal{O}_{i}^{(1/2)}$,
then we have for the matrix elements, 
\begin{eqnarray*}
	\mathcal{M}_{+} & = & 
	\left\langle \pi^{+}\left|\mathcal{O}_{i}^{(3/2)}
	\right|K^{+}\right\rangle 
	+\left\langle \pi^{+}\left|\mathcal{O}_{i}^{(1/2)}\right|K^{+}
	\right\rangle, \\
	\mathcal{M}_{0} & = & 
	\left\langle \pi^{0}\left|\mathcal{O}_{i}^{(3/2)}
	\right|K^{0}\right\rangle +\left\langle \pi^{0}
	\left|\mathcal{O}_{i}^{(1/2)}\right|K^{0}\right\rangle .
\end{eqnarray*}
Given the relevant Clebsch-Gordon coefficients,
\begin{eqnarray}
	\frac{\left\langle \pi^{+}\left|\mathcal{O}_{i}^{(3/2)}
	\right|K^{+}\right\rangle }{\left\langle \pi^{0}\left|
	\mathcal{O}_{i}^{(3/2)}\right|K^{0}\right\rangle } & = &
	\frac{\sqrt{2}}{2},\\
	\frac{\left\langle \pi^{+}\left|\mathcal{O}_{i}^{(1/2)}\right|K^{+}\right\rangle }{\left\langle \pi^{0}\left|\mathcal{O}_{i}^{(1/2)}\right|K^{0}\right\rangle }
	&=&
	-\sqrt{2},
	\label{eq:solve isospin decomp}
\end{eqnarray}
we obtain the result,
\begin{align}
	\left\langle \pi^{+}\left|\mathcal{O}_{i}^{(3/2)}\right|K^{+}
	\right\rangle  & =\frac{1}{3}\left(\mathcal{M}_{+}+\sqrt{2}
	\mathcal{M}_{0}\right)\nonumber \\
	\left\langle \pi^{+}\left|\mathcal{O}_{i}^{(1/2)}\right|K^{+}
	\right\rangle  & =\frac{1}{3}\left(2\mathcal{M}_{+}-\sqrt{2}
	\mathcal{M}_{0}\right)\ .
\label{eq:Isospin decomp}
\end{align}

\section{Loop Functions and Residues\label{sec:Loop-Functions}}\label{sec:loopfunc}

The following loop functions are used throughout this work, and they are regulated using dimensional regularization in the modified $\overline{\textrm{MS}}$ scheme.
For single-pole mesonic loops, we need
\begin{equation}
	\ell\left(m^{2}\right)
	=\left[\lim_{d\to4}\int\frac{d^{d}p}{\left(2\pi\right)^{d}}
	\frac{i}{p^{2}-m^{2}+i\epsilon}\right]_{\mathrm{reg}}
	=\frac{1}{16\pi^{2}}m^{2}\ln\left(\frac{m^{2}}{\mu^{2}}\right)\ ,
	\label{eq:LoopFunc}
\end{equation}
(cf. $f^2A\left(m^{2}\right)$ in Ref.~\cite{Laiho:2002jq, Laiho:2003uy}). We also need the double pole expression
\begin{eqnarray}
	\tilde{\ell}\left(m^{2}\right) & = &
	-\frac{\partial}{\partial m^{2}}\ell\left(m^{2}\right)
	\nonumber \\
	& = & -\int\frac{d^{d}p}{\left(2\pi\right)^{d}}
	\frac{i}{\left(p^{2}-m^{2}\right)^{2}}\ ,
	\label{eq:LoopTilde}
\end{eqnarray}
where the minus sign is chosen to be consistent with the form of Euclidean
$\tilde{\ell}\left(m^{2}\right)$ in Refs.~\cite{Aubin:2003mg,Aubin:2003uc}.\footnote{Note, however, that our definitions of $\ell$ and $\tilde\ell$ differ from Refs.~\cite{Aubin:2003mg,Aubin:2003uc} by a factor of $1/16\pi^2$.} Further,
we will sometimes need
\begin{eqnarray}
	\tilde{\tilde{\ell}}\left(m^{2}\right) & =&
	\frac{\partial}{\partial m^{2}}\tilde{\ell}\left(m^{2}\right).
\label{eq:LoopTildeTilde}
\end{eqnarray}

The two-point loop-function encountered in loops with strong-weak
vertices and only a single pole is defined as:
\begin{eqnarray}
	\beta\left(q^{2},m_{1}^{2},m_{2}^{2}\right) 
	& = & \left[ 
	i\int\frac{d^{d}p} {\left( 2\pi \right)^{d}}
	\frac{1}{\left( p^{2}-m_{1}^{2} \right) 
	\left( \left( p+q \right)^{2} - m_{2}^{2} \right)}
   \right]_{\mathrm{reg}}							
   \nonumber \\
	& = &  \frac{1}{ \left( 4 \pi \right)^2 } 
	\int_0^1 dx 
	\left\{ 1 + \ln \left[ -x (1-x) q^2 + (1-x) 
	m_1^2 + x m_2^2 \right] - \ln \left( \mu^2 \right) \right\}. \nonumber \\ &&
	\label{eq:BLoopFunc}
\end{eqnarray}
Note that we always have $q^{2}=\left(m_{xz}-m_{X}\right)^{2}$
for $K\to\pi$ amplitudes.  This function is proportional to the $B_{0}$
function defined in Eq (A2) of \cite{Laiho:2002jq}. Similar loops with double-poles require
\begin{equation}
	\tilde{\beta}\left(q^{2},m_{1}^{2},m_{2}^{2}\right)
	=\frac{\partial}{\partial\left(m_{1}^{2}\right)}
	\beta\left(q^{2},m_{1}^{2},m_{2}^{2}\right)
	\label{eq:beta_tilde_LoopFunc}\ .
\end{equation}

\bigskip

To simplify the expressions, we use the
notation for the residues arising from disconnected meson propagators,
\begin{align}
	R_{x}\left(m_{a}\right) 
	& =\frac{\left(m_{x}^{2}-m_{D}^{2}\right)
	\left(m_{x}^{2}-m_{S}^{2}\right)}{m_{x}^{2}-m_{a}^{2}}\ ,
	\label{eq:residue-1}\\
	R_{x}\left(m_{a},m_{b}\right) 
	& =\frac{\left(m_{x}^{2}-m_{D}^{2}\right)
	\left(m_{x}^{2}-m_{S}^{2}\right)}
	{\left(m_{x}^{2}-m_{a}^{2}\right)
	\left(m_{x}^{2}-m_{b}^{2}\right)} \ .
\label{eq:residue-2}
\end{align}

\section{One-loop wavefunction and mass renormalizations}\label{sec:renorms}

The necessary wavefunction renormalizations needed for the one-loop amplitudes are given in the 2+1 flavor case by
\begin{eqnarray}\label{eq:Zxz}
  \delta Z_{xz} & =&
  	-2 \frac{\Delta f_{xz}}{f}
  	+\frac{4}{3}\left(\frac{\Delta f_{xz}}{f}\right)_{\rm logs},
\end{eqnarray}
\begin{eqnarray}\label{eq:Zx}
  \delta Z_{X} & =&
  	-2 \frac{\Delta f_{X}}{f}
  	+\frac{4}{3}\left(\frac{\Delta f_{X}}{f}\right)_{\rm logs},
\end{eqnarray}
where we have separated the terms in this way because the first term on the right-hand side of each of these equations is the one-loop correction to the bare decay constant $f$ appearing in the tree-level expression for a given weak matrix element.  This first term contains both NLO logarithmic corrections and Gasser-Leutwyler constants.  It may be useful in chiral fits to lattice data to absorb this correction to the decay constant into the tree-level expression, and the above formulas make this convenient.  The second term is proportional to the logarithmic corrections to the decay constant alone, without the Gasser-Leutwyler constants,
\begin{eqnarray}
 \left( \frac{\Delta f_{xz}}{f}\right)_{\rm logs} & =&
  \frac{1}{2 f^2}\Biggl[
  	-\left[
  		2\ell\left(m_{xd}^2\right)  +2\ell\left(m_{zd}^2\right)
  		+\ell\left(m_{xs}^2\right)  +\ell\left(m_{zs}^2\right)
  	\right]\nonumber \\* && \qquad
  	+\frac{1}{3}\Biggl(
    		 \frac{\partial R_{X}\left(m_\eta\right)}{\partial m_{X}^2}
  			 \ell(m_X^2)-R_{X}(m_\eta)\tilde{\ell}(m^2_X)
    		+ \frac{\partial R_{\eta} \left(m_X\right)}{\partial m_{X}^2}
  			\ell(m_\eta^2)
  								\nonumber \\* && \qquad
    		+ \frac{\partial R_{Z}  \left(m_\eta\right)}{\partial m_{Z}^2} 
  			 \ell(m_{Z}^2) -R_Z(m_\eta)\tilde{\ell}(m_Z^2)
		+ \frac{\partial R_{\eta}  \left(m_Z\right)}{\partial m_{Z}^2} 
  			 \ell(m_\eta^2)
  								\nonumber \\* && \qquad
  		- 2 R_{X} \left(m_Z,m_\eta\right) \ell(m_X^2)
  		- 2 R_{Z} \left(m_X,m_\eta\right) \ell(m_Z^2)  \nonumber \\* && \qquad
  		- 2 R_{\eta} \left(m_X,m_Z\right) \ell(m_\eta^2)  
	\Biggr) 
  \Biggr],        					
  \end{eqnarray}
  so that \cite{Sharpe:2000bc},
  \begin{eqnarray}
  \frac{\Delta f_{xz}}{f}& =&  \left( \frac{\Delta f_{xz}}{f}\right)_{\rm logs}
  + \frac{8}{f^2}L_4 \left(2m_D^2 + m_S^2\right)
  + \frac{8}{f^2}L_5m_{xz}^2\ .
\end{eqnarray}
For the degenerate mass case, these reduce to
\begin{eqnarray}
 \left( \frac{\Delta f_{X}}{f}\right)_{\rm logs} & =&
  \frac{1}{f^2}\left[
  	-2\ell\left(m_{xd}^2\right)  
  		-\ell\left(m_{xs}^2\right)  
  \right],        					
  \end{eqnarray}
  \begin{eqnarray}
  \frac{\Delta f_{X}}{f}& =&  \left( \frac{\Delta f_{X}}{f}\right)_{\rm logs}
  + \frac{8}{f^2}L_4 \left(2m_D^2 + m_S^2\right)
  + \frac{8}{f^2}L_5m_{X}^2\ .
\end{eqnarray}
Additionally, we need the one-loop corrections to the meson masses squared  \cite{Sharpe:2000bc},
\begin{eqnarray}
  \frac{(\Delta m_{xz})^2}
  {m_{xz}^2 }&=& 
  \frac{2}{3f^2}\Big( 
   R_{X}(m_Z,m_\eta)\;  \ell(m^2_{X})
  +R_{Z}(m_X,m_\eta)\;  \ell(m^2_{Z}) 
  + R_{\eta}(m_X,m_Z)\;  \ell(m^2_{\eta})\Big) 
   \nonumber \\*
   &&+\frac{16}{f^2}\left(2L_8-L_5\right)m_{xz}^2
   +\frac{16}{f^2}\left(2L_6-L_4\right)
   \left(2m_D^2 + m_S^2\right),
\end{eqnarray}
\begin{eqnarray}
  \frac{(\Delta m_{X})^2}
  {m_{X}^2 }&=& 
  \frac{2}{3f^2}\Big( 
   -R_{X}(m_\eta)\;  \tilde{\ell}(m^2_{X})
  +\frac{\partial R_X(m_\eta)}{\partial m^2_X}\;  \ell(m^2_{X}) 
  + R_{\eta}(m_X,m_X)\;  \ell(m^2_{\eta})\Big) 
   \nonumber \\*
   &&+\frac{16}{f^2}\left(2L_8-L_5\right)m_{X}^2
   +\frac{16}{f^2}\left(2L_6-L_4\right)
   \left(2m_D^2 + m_S^2\right).
\end{eqnarray}

\section{Logarithmic Contribution to $(3,\bar{3})$ $K\to\pi$ Matrix Elements \label{sub:logterms 33}}

The logarithmic contribution to the $(3,\overline{3})$, $K\to\pi$ matrix element for the 2+1 non-degenerate case is,

\begin{align*}
	\left\langle \pi^{+}\left|\Theta^{\left(3,\bar{3}\right)}
	\right|K^{+}\right\rangle_{\rm logs} =&	
	-\frac{2}{f^{2}}\alpha^{(3,\bar{3})}\left[
	\frac{1}{2}\delta Z_{xz}+\frac{1}{2}\delta Z_{X}\right]
	\nonumber \\
	 & + \frac{2}{9}\frac{\alpha^{(3,\bar{3})}}{f ^{4} }
	 \biggl\{-\frac{9m_{X}(2\ell(m_{xd}^{2})  
	 +\ell(m_{xs}^{2}))}{m_{xz}-m_{X}} \nonumber\\	 
 	& +\biggl[-2  +\frac{3m_{X}}{m_{xz}-m_{X}}  
	+\biggl(2  -\frac{3m_{X}}{m_{xz}-m_{X}}\biggr)R_{\eta}(m_{X},m_{X})  
	\nonumber\\
 	&-2R_{X}(m_{Z},m_{\eta})\biggr]\ell(m_{X}^{2}) \nonumber\\
 	& +\biggl[-2  -\frac{3m_{X}}{m_{xz}-m_{X}}  
	+\biggl(2  +\frac{3m_{X}}{m_{xz}-m_{X}}\biggr)R_{\eta}(m_{Z},m_{Z})  
	\nonumber\\
 	&-2R_{Z}(m_{X},m_{\eta})\biggr]\ell(m_{Z}^{2}) \nonumber\\
 	& +6\biggl[2  +\frac{3m_{X}}{m_{xz}-m_{X}}\biggr]\ell(m_{zd}^{2})  
	+3\biggl[2  +\frac{3m_{X}}{m_{xz}-m_{X}}\biggr]\ell(m_{zs}^{2}) \nonumber\\
 	& +\biggl[\biggl(-2  +\frac{3m_{X}}{m_{xz}-m_{X}}\biggr)R_{\eta}(m_{X},m_{X})  
	-2R_{\eta}(m_{X},m_{Z})  \nonumber\\
	& -\biggl(2  +\frac{3m_{X}}{m_{xz}-m_{X}}\biggr)R_{\eta}(m_{Z},m_{Z})\biggr]
	\ell(m_{\eta}^{2})
\end{align*}

\begin{align}
 & -\biggl[\biggl(2  +\frac{3m_{X}}{m_{xz}-m_{X}}\biggr)R_{X}(m_{\eta})  
 +2\biggl(2m_{xz}^{2}  +m_{X}^{2}\biggr)R_{X}(m_{Z},m_{\eta})\biggr]
 \beta(q^{2},m_{xz}^{2},m_{X}^{2}) \nonumber\\
 & +\biggl[2m_{Z}^{2}  -4m_{xz}^{2}  
 +\frac{3m_{X}(m_{Z}^{2}  -m_{X}^{2})}{m_{xz}-m_{X}}  -4m_{X}^{2}  
 +\biggl(2  +\frac{3m_{X}}{m_{xz}-m_{X}}\biggr)R_{Z}(m_{\eta}) \nonumber\\
 & +\biggl(-2m_{Z}^{2}  +4m_{xz}^{2}  +\frac{3m_{X}(-m_{Z}^{2}  
 +m_{X}^{2})}{m_{xz}-m_{X}}  +4m_{X}^{2}\biggr)R_{\eta}(m_{Z},m_{Z}) \nonumber\\
 & -2\biggl(2m_{xz}^{2}  +m_{X}^{2}\biggr)R_{Z}(m_{X},m_{\eta})\biggr]
 \beta(q^{2},m_{Z}^{2},m_{xz}^{2}) \nonumber\\
 & +6\biggl[-2m_{zd}^{2}  +2m_{xd}^{2}  +2m_{xz}^{2}  
 +3m_{X}\biggl(\frac{-m_{zd}^{2}  +m_{xd}^{2}}{m_{xz}-m_{X}}  -m_{xz}\biggr)
 +m_{X}^{2}\biggr]\beta(q^{2},m_{zd}^{2},m_{xd}^{2}) \nonumber\\
 & +3\biggl[-2m_{zs}^{2}  +2m_{xs}^{2}  +2m_{xz}^{2}  
 +3m_{X}\biggl(\frac{-m_{zs}^{2}  +m_{xs}^{2}}{m_{xz}-m_{X}}  -m_{xz}\biggr)
 +m_{X}^{2}\biggr]\beta(q^{2},m_{zs}^{2},m_{xs}^{2}) \nonumber\\
 & +\biggl[\biggl(-2m_{\eta}^{2}  +\frac{3m_{X}(-m_{\eta}^{2}  +m_{X}^{2})}
 {m_{xz}-m_{X}}  +2m_{X}^{2}\biggr)R_{\eta}(m_{X},m_{X}) \nonumber\\
 & -2\biggl(2m_{xz}^{2}  +m_{X}^{2}\biggr)R_{\eta}(m_{X},m_{Z}) \nonumber\\
 & +\biggl(2m_{\eta}^{2}  -4m_{xz}^{2}  +\frac{3m_{X}(m_{\eta}^{2}  
 -m_{X}^{2})}{m_{xz}-m_{X}}  -4m_{X}^{2}\biggr)R_{\eta}(m_{Z},m_{Z})\biggr]
 \beta(q^{2},m_{\eta}^{2},m_{xz}^{2})    \nonumber\\
 & + \biggl[2  -\frac{3m_{X}}{m_{xz}-m_{X}}\biggr]R_{X}(m_{\eta})\tilde{\ell}(m_{X}^{2})
 +\biggl[2  +\frac{3m_{X}}{m_{xz}-m_{X}}\biggr]R_{Z}(m_{\eta})\tilde{\ell}(m_{Z}^{2}) \nonumber\\
 & + \biggl[2m_{Z}^{2}  -4m_{xz}^{2}  +\frac{3m_{X}(m_{Z}^{2}  -m_{X}^{2})}{m_{xz}-m_{X}}
 -4m_{X}^{2}\biggr]
 R_{Z}(m_{\eta})\tilde{\beta}(q^{2},m_{Z}^{2},m_{xz}^{2}) \biggr\} 	\label{eq:O33 Kpi logs2}
\end{align}

\section{Logarithmic Contribution to (8,8) $K \to \pi$ Matrix Elements \label{sub:logterms 88}}

The logarithmic contribution to the (8,8), $\Delta I=3/2$, $K\to\pi$ matrix element in the 2+1 non-degenerate case is
\begin{eqnarray}
	\left\langle \pi^{+}\left|\mathcal{O}^{\left(8,8\right)(3/2)}
	\right|K^{+}\right\rangle _{Q_{1},{\rm logs}}
	& = & \frac{4\alpha_{88}}{f^{2}}
	\left(\frac{1}{2}\delta Z_{X}+\frac{1}{2}
	\delta Z_{xz}\right)
	\nonumber \\&&{}
	+ \frac{8}{9}\frac{\alpha_{88}}{f^{4}}\biggl\{
	\biggl[1-2R_{X}(m_{Z},m_{\eta})-R_{\eta}(m_{X},m_{X})
	\biggr]\ell(m_{X}^{2})
	\nonumber \\&&{}
 	+\biggl[1-2R_{Z}(m_{X},m_{\eta})
	-R_{\eta}(m_{Z},m_{Z})\biggr]\ell(m_{Z}^{2})
	\nonumber \\&&{}
	-18\ell(m_{xd}^{2})-9\ell(m_{xs}^{2})
	-6\ell(m_{zd}^{2})-3\ell(m_{zs}^{2})
	\nonumber \\&&{}
 	+\biggl[R_{\eta}(m_{Z},m_{Z})-2R_{\eta}(m_{X},m_{Z})
	+R_{\eta}(m_{X},m_{X})\biggr]\ell(m_{\eta}^{2})
	\nonumber \\&&{}
	-R_{X}(m_{\eta})\tilde{\ell}(m_{X}^{2})
	-R_{Z}(m_{\eta})\tilde{\ell}(m_{Z}^{2})
	\nonumber \\&&{}
	-9m_{xz}m_{X}
	\beta(q^{2},m_{xz}^{2},m_{X}^{2})\biggr\}.
\label{eq:O88 Kpi Q1 3/2 logs}
\end{eqnarray}
The logarithmic contribution to the (8,8), $\Delta I=1/2$, $K\to\pi$ matrix element in the 2+1 non-degenerate case is
\begin{eqnarray}
 	\left\langle \pi^{+}\left|\mathcal{O}^{\left(8,8\right)(1/2)}
	\right|K^{+}\right\rangle _{Q_{1},{\rm logs}}
	& = & \frac{8\alpha_{88}}{f^{2}}
	\left(\frac{1}{2}\delta Z_{X}+\frac{1}{2}
	\delta Z_{xz}\right)
	\nonumber \\&&{}
	+ \frac{2}{9}\frac{\alpha_{88}}{f^{4}}\biggl\{8
	\biggl[1-2R_{X}(m_{Z},m_{\eta})-R_{\eta}(m_{X},m_{X})\biggr]
	\ell(m_{X}^{2})
	\nonumber \\&&{}
	+8\biggl[1-2R_{Z}(m_{X},m_{\eta})-R_{\eta}(m_{Z},m_{Z})\biggr]
	\ell(m_{Z}^{2})
	\nonumber \\&&{}
 	+18\frac{4m_{X} - 3m_{xz}}{m_{xz}-m_{X}}
	\left[2\ell(m_{xd}^{2}) + \ell(m_{xs}^{2})\right]
	\nonumber \\&&{}
 	+6\frac{4m_{X} - 7m_{xz}}{m_{xz}-m_{X}}
	\left[2\ell(m_{zd}^{2}) + \ell(m_{zs}^{2})\right]
	\nonumber \\&&{}
	+8\biggl[R_{\eta}(m_{Z},m_{Z})-2R_{\eta}(m_{X},m_{Z})
	+R_{\eta}(m_{X},m_{X})\biggr]\ell(m_{\eta}^{2})
	\nonumber \\&&{}
	-8R_{X}(m_{\eta})\tilde{\ell}(m_{X}^{2})
	-8R_{Z}(m_{\eta})\tilde{\ell}(m_{Z}^{2})
	\nonumber \\&&{}
	+36m_{xz}m_{X}\beta(q^{2},m_{xz}^{2},m_{X}^{2})
	\nonumber \\&&{}
 	+36m_Xm_{xz}
	\left(2\beta(q^{2},m_{zd}^{2},m_{xd}^{2})+
	\beta(q^{2},m_{zs}^{2},m_{xs}^{2})
	\right)
	\biggr\}.  
\label{eq:O88 Kpi Q1 1/2 logs}                       
\end{eqnarray}

For completeness we include the chiral corrections to $K\to\pi\pi$ at physical kinematics for the electro-weak penguin operators.
In the full theory at physical kinematics, the logarithmic contribution to the (8,8), $\Delta I=3/2$, $K\to\pi\pi$ amplitude is
\bea \bra\pi^+\pi^- | {\cal O}^{(8,8),(3/2)} | K^0\ket_{\rm logs} & = &
-4i \frac{\alpha_{88}}{f_Kf^2_\pi f^2} \left[ \left(
\frac{5m^4_K}{4m^2_\pi} -2m^2_K\right) \beta(m^2_\pi, m^2_K, m^2_\pi)\right. \nonumber \\
& & +(m^2_K-2m^2_\pi) \beta(m^2_K, m^2_\pi, m^2_\pi) \nonumber \\
& & + \frac{m^4_K}{4m^2_\pi} \beta(m^2_\pi, m^2_K,
m^2_\eta)\nonumber
- \left(4+ \frac{m^2_K}{2m^2_\pi}\right) \ell(m^2_K)\nonumber \\
& & + \left( \frac{5m^2_K}{4m^2_\pi} -8 \right )
\ell(m^2_\pi)\left. - \frac{3m^2_K}{4m^2_\pi} \ell(m^2_\eta)\right],
\eea
and the logarithmic contribution to the (8,8), $\Delta I=1/2$, $K\to\pi\pi$ amplitude is
\bea \bra\pi^+\pi^- | {\cal O}^{(8,8),(1/2)} | K^0\ket_{\rm logs} & = &
-8i \frac{\alpha_{88}}{f_Kf^2_\pi f^2} \left[ \left(
\frac{m^4_K}{2m^2_\pi} -2m^2_K\right) \beta(m^2_\pi, m^2_K, m^2_\pi)\right. \nonumber \\
& &+\frac{3}{4}m^2_K \beta(m^2_K,m^2_K,m^2_K) +(m^2_\pi-2m^2_K) \beta(m^2_K, m^2_\pi, m^2_\pi) \nonumber \\
& & + \frac{m^4_K}{4m^2_\pi} \beta(m^2_\pi, m^2_K,
m^2_\eta)\nonumber
+\frac{1}{4} \left(\frac{m^2_K}{m^2_\pi}-22 \right) \ell(m^2_K)\nonumber \\
& & + \frac{1}{4}\left( \frac{2m^2_K}{m^2_\pi} -26 \right )
\ell(m^2_\pi)\left. - \frac{3m^2_K}{4m^2_\pi} \ell(m^2_\eta)\right].
\eea

\section{Logarithmic Contribution to (27,1) $K\to\pi$ Matrix Elements \label{sub:logterms 271}}

The logarithmic contribution to the (27,1), $\Delta I=3/2$, $K\to\pi$ matrix element in the 2+1 non-degenerate case is
\begin{eqnarray}
 	\left\langle \pi^{+}\left|\mathcal{O}^{\left(27,1\right)(3/2)}
	\right|K^{+}\right\rangle_{\rm logs} 
	& = & -\frac{4\alpha_{27}}{f^{2}}m_{X}m_{xz}\left[\frac{1}{2}
	\delta Z_{X}+\frac{1}{2}\delta Z_{xz}+\frac{1}{2}\left(
	\frac{\delta m_{X}^{2}}{m_{X}^{2}}
	+\frac{\delta m_{xz}^{2}}{m_{xz}^{2}}\right)\right]
	\nonumber \\&&{}
	+ \frac{8}{9}\frac{\alpha_{27}}{f^{4}}m_{X}m_{xz}\biggl\{
	\biggl[8+R_{\eta}(m_{X},m_{X})+2R_{X}(m_{Z},m_{\eta})\biggr]
	\ell(m_{X}^{2})
	\nonumber \\&&{}
	+\biggl[-1+R_{\eta}(m_{Z},m_{Z})+2R_{Z}(m_{X},m_{\eta})
	\biggr]\ell(m_{Z}^{2})
	\nonumber \\&&{}
	+9\ell(m_{xz}^{2})+18\ell(m_{xd}^{2})+9\ell(m_{xs}^{2})
	+6\ell(m_{zd}^{2})+3\ell(m_{zs}^{2})
	\nonumber \\&&{}
	+\biggl[-R_{\eta}(m_{X},m_{X})+2R_{\eta}(m_{X},m_{Z})-R_{\eta}
	(m_{Z},m_{Z})\biggr]\ell(m_{\eta}^{2})
	\nonumber \\&&{}
	+R_{X}(m_{\eta})\tilde{\ell}(m_{X}^{2})+R_{Z}(m_{\eta})
	\tilde{\ell}(m_{Z}^{2})+9m_{X}m_{xz}
	\beta(q^{2},m_{xz}^{2},m_{X}^{2})\biggr\} . \nonumber\\
	\label{eq:O27 Kpi 3/2 logs}
\end{eqnarray}
The logarithmic contribution to the (27,1), $\Delta I=1/2$, $K\to\pi$ matrix element in the 2+1 non-degenerate case is
\begin{eqnarray*}
 	\left\langle \pi^{+}\left|\mathcal{O}^{\left(27,1\right)(1/2)}
	\right|K^{+}\right\rangle_{\rm logs} 
	& = & 
	-\frac{4\alpha_{27}}{f^{2}}m_{xz}m_{X}\left[
	\frac{1}{2}\delta Z_{X}+\frac{1}{2}\delta Z_{xz}
	+\frac{1}{2}\left(\frac{\delta m_{X}^{2}}{m_{X}^{2}}
	+\frac{\delta m_{xz}^{2}}{m_{xz}^{2}}\right)\right]
	\\&&{}
	+\frac{2}{9}\frac{\alpha_{27}}{f^{4}}\biggl\{
	12m_{xz}m_{X}\left(
	6\ell(m_{xd}^{2}) + 3\ell(m_{xs}^{2}) + 2\ell(m_{zd}^{2})
	+\ell(m_{zs}^{2})\right)
	\\&&{}
	+\biggl[\biggl(12m_{\eta}^{2}-4m_{xz}m_{X}+6m_{X}^{2}
	\\&&{}
	+\frac{9m_{X}(m_{\eta}^{2}+m_{X}^{2})}{m_{xz}-m_{X}}\biggr)
	R_{\eta}(m_{X},m_{X})
	\\&&{}
 	-2\biggl(12m_{\eta}^{2} + 6m_{xz}^{2} - 4m_{xz}m_{X} 
	+ 9m_{X}^{2} 
	\\&&{}
	+ \frac{9m_{X}(m_{\eta}^{2}+m_{X}^{2})}{m_{xz}-m_{X}}\biggr)
	R_{\eta}(m_{X},m_{Z})
	\\&&{}
	+\biggl(12m_{\eta}^{2}+12m_{xz}^{2}-4m_{xz}m_{X}+12m_{X}^{2}
	\\&&{}
	+\frac{9m_{X}(m_{\eta}^{2}
	+m_{X}^{2})}{m_{xz}-m_{X}}\biggr)
	R_{\eta}(m_{Z},m_{Z})\biggr]\ell(m_{\eta}^{2})\biggr\}\\
	&&{}+
	\frac{4\alpha_{27} m_{xz}}{f^4}
	\left[ 5m_X + \frac{6m^2_{xz}}{m_{xz}-m_X} \right] \ell(m^2_{xz})
\end{eqnarray*}
\begin{eqnarray*}
	&&{}+ \frac{2}{9}\frac{\alpha_{27}}{f^{4}}\biggl\{
	\biggl[
	\frac{4 m_X m_{xz} (10 m_X - 19 m_{xz} )}{m_{xz} - m_{X}}
	+3\frac{4m_{xz} - m_{X}}{m_{xz}-m_{X}}
	R_{X}(m_{\eta})
	\\&&{}
	+2
	\frac{m_X m_{xz}(2m_{xz} - 11m_X)}{m_{xz} - m_X}
	R_{\eta}(m_{X},m_{X})
	\\&&{}
	+2\biggl(-6m_{xz}^{2}+4m_{xz}m_{X}-21m_{X}^{2}
	-\frac{18m_{X}^{3}}{m_{xz}-m_{X}}\biggr)
	R_{X}(m_{Z},m_{\eta})\biggr]\ell(m_{X}^{2})
	\\&&{}
	+\biggl[-72m_{xz}^{2} - 4m_{xz}m_{X} 
	-\frac{36m_{X}m_{xz}^{2}}{m_{xz}-m_{X}}
	+3\frac{4m_{xz} - m_{X}}{m_{xz}-m_{X}}
	R_{Z}(m_{\eta})
	\\&&{}
	+\biggl(
	-36m_{xz}^{2} + 4m_{xz}m_{X}
	-\frac{18m_{X}m_{xz}^{2}}{m_{xz}-m_{X}}\biggr)
	R_{\eta}(m_{Z},m_{Z})
	\\&&{}
	+2\biggl(
	-15m_{Z}^{2} + 4m_{xz}m_{X} - 12 m_{X}^{2}
	-\frac{18m_{X}m_{xz}^{2}}{m_{xz}-m_{X}}\biggr)
	R_{Z}(m_{X},m_{\eta})\biggr]\ell(m_{Z}^{2})
	\\&&{}
	+\frac{2m_{X}m_{xz}(2m_{xz} - 11m_X)}{m_{xz} - m_X}
	R_{X}(m_{\eta})\tilde{\ell}(m_{X}^{2})
	\\&&{}
	+\biggl[4m_{xz}m_{X} - 36m_{xz}^{2}
	-\frac{18m_{X}m_{xz}^{2}}{m_{xz}-m_{X}}\biggr]
	R_{Z}(m_{\eta})\tilde{\ell}(m_{Z}^{2})\biggr\}
\end{eqnarray*}
\begin{eqnarray*}
	&&+ \frac{2}{3}\frac{\alpha_{27}}{f^{4}}\biggl\{
	\biggl[-2m_{\eta}^{4}-4m_{\eta}^{2}m_{xz}m_{X}
	-2m_{\eta}^{2}m_{X}^{2}+4m_{xz}m_{X}^{3}+4m_{X}^{4}
	\\&&{}
	+\frac{3m_{X}(m_{X}^{4} - m_{\eta}^{4})}{m_{xz}-m_{X}}\biggr]
	R_{\eta}(m_{X},m_{X})
	\\&&{}
	+2\biggl[2m_{\eta}^{4}-2m_{\eta}^{2}m_{xz}^{2}+4m_{\eta}^{2}
	m_{xz}m_{X}-4m_{xz}^{3}m_{X}+m_{\eta}^{2}m_{X}^{2}+2m_{xz}^{2}
	m_{X}^{2}
	\\&&{}
	-6m_{xz}m_{X}^{3}-3m_{X}^{4}+\frac{3m_{X}(m_{\eta}^{4} - m_{X}^{4})}
	{m_{xz}-m_{X}}\biggr]R_{\eta}(m_{X},m_{Z})
	\\&&{}
	+\biggl[-2m_{\eta}^{4}+4m_{\eta}^{2}m_{xz}^{2}-4m_{\eta}^{2}
	m_{xz}m_{X}+8m_{xz}^{3}m_{X}-4m_{xz}^{2}m_{X}^{2}+8m_{xz}
	m_{X}^{3}
	\\&&{}
	+2m_{X}^{4}+\frac{3m_{X}(m_{X}^{4}-m_{\eta}^{4})}{m_{xz}-m_{X}}
	\biggr]R_{\eta}(m_{Z},m_{Z})\biggr\}
	\beta(q^{2},m_{\eta}^{2},m_{xz}^{2})
\end{eqnarray*}
\begin{eqnarray}
	&&+ \frac{2}{3}\frac{\alpha_{27}}{f^{4}}\biggl\{-2
	m_X m_{xz} 
	\biggl[12 m_{xz} m_X
	-\frac{m_X + 2 m_{xz}}{m_X - m_{xz}}
	R_{X}(m_{\eta}) 
	\nonumber \\&&{}
	+2\biggl(2m_{xz}^2 + m_X^2\biggr)
	R_{X}(m_{Z},m_{\eta})\biggr]
	\beta(q^{2},m_{xz}^{2},m_{X}^{2}) 
	\nonumber \\&&{}
	+\biggl[
	6m_X\left(
	-4 m_X^2 m_{xz}+\frac{m_{Z}^{4}-m_{X}^{4}}{m_{xz}-m_{X}}\right)
	+2\biggl(2m_{xz}^{2}-2m_{Z}^{2}-2m_{xz}m_{X}
	\nonumber\\&&{}
	-\frac{3m_{Z}^{2}m_{X}}{m_{xz}-m_{X}}\biggr)
	R_{Z}(m_{\eta})
	 \nonumber \\&&{}
	+3m_X\biggl(
	-4m_X^2 m_{xz}
	+\frac{m_{Z}^{4}-m_{X}^4}{m_{xz}-m_{X}}
	\biggr)R_{\eta}(m_{Z},m_{Z}) 
	\nonumber \\&&{}
	+2\biggl(
	m_Z^4-2m_X^3(m_X + 4m_{xz}) + m_X m_Z^2 (m_X +2 m_{xz})
	\nonumber \\&&{}
	+\frac{3m_{X}(m_{Z}^{4}-m_{X}^{4})}{m_{xz}-m_{X}}\biggr)
	R_{Z}(m_{X},m_{\eta})\biggr]
	\beta(q^{2},m_{xz}^{2},m_{Z}^{2})
	 \nonumber \\&&{}
	+ 3 m_X\biggl[
	4 m_X^2 m_{xz}
	+\frac{m_{X}^{4}-m_{Z}^{4}}{m_{xz}-m_{X}}\biggr]
	R_{Z}(m_{\eta})\tilde{\beta}(q^{2},m_{Z}^{2},m_{xz}^{2})
	\biggr\}.
	\label{eq:O27 Kpi 1/2 logs}
\end{eqnarray}

\section{Logarithmic Contribution to (8,1) $K\to 0$ and $K\to\pi$ Matrix Elements \label{sub:logterms 81}}

For the (8,1) case, we separate the logarithm terms which are proportional to $\alpha_1$ from those proportional to $\alpha_2$. 

For $K\to 0$, we have 
\begin{equation}
 	\left\langle 0\left|\mathcal{O}^{\left(8,1\right)}
	\right|K^{0}\right\rangle_{\rm logs}
	=
 	\left\langle 0\left|\mathcal{O}^{\left(8,1\right)}
	\right|K^{0}\right\rangle_{\rm logs}^{(1)}
	+
 	\left\langle 0\left|\mathcal{O}^{\left(8,1\right)}
	\right|K^{0}\right\rangle_{\rm logs}^{(2)}\ ,
\end{equation}
where the superscripts refer to the terms proportional to $\alpha_{1,2}$, and 
\begin{eqnarray}
 	\left\langle 0\left|\mathcal{O}^{\left(8,1\right)}
	\right|K^{0}\right\rangle_{\rm logs}^{(1)}
	& = &\frac{4i}{3f^{3}}\alpha_{1}\biggl\{
	\biggl[m_{X}^{2}+R_{X}(m_{\eta}) - 
	m_{X}^{2}R_{\eta}(m_{X},m_{X})\biggr]\ell(m_{X}^{2})
	\nonumber \\&&{}
 	+\biggl[-m_{Z}^{2}-R_{Z}(m_{\eta})+m_{Z}^{2}R_{\eta}
	(m_{Z},m_{Z})\biggr]\ell(m_{Z}^{2})
	\nonumber \\&&{}
 	+ m_{\eta}^{2}\biggl[R_{\eta}(m_{X},m_{X}) 
	- R_{\eta}(m_{Z},m_{Z})\biggr]\ell(m_{\eta}^{2})
  	\nonumber \\&&{}
 	-6m_{xd}^{2}\ell(m_{xd}^{2})-3m_{xs}^{2}\ell(m_{xs}^{2})
	+6m_{zd}^{2}\ell(m_{zd}^{2})+3m_{zs}^{2}\ell(m_{zs}^{2})
 	\nonumber \\&&{}
 	-m_{X}^{2}R_{X}(m_{\eta})\tilde{\ell}(m_{X}^{2})
	+m_{Z}^{2}R_{Z}(m_{\eta})\tilde{\ell}(m_{Z}^{2})\biggr\} ,                                  \label{eq:O81 Kvac logs1}
\end{eqnarray}
\begin{eqnarray}
 	\left\langle 0\left|\mathcal{O}^{\left(8,1\right)}
	\right|K^{0}\right\rangle_{\rm logs}^{(2)}
	& = & \frac{4i}{f}\alpha_{2}(m_{xz}^{2}-m_{X}^{2})
	\left[\frac{1}{2}\delta Z_{xz}\right]
	\nonumber \\&&{}
	+ \frac{8i}{9}\frac{\alpha_{2}}{f^{3}}(m_{xz}^{2}-m_{X}^{2})
	\biggl\{\biggl[1+R_{X}(m_{\eta},m_{Z})-R_{\eta}(m_{X},m_{X})
	\biggr]\ell(m_{X}^{2})
	\nonumber \\&&{}
	+\biggl[1+R_{Z}(m_{\eta},m_{X})-R_{\eta}(m_{Z},m_{Z})\biggr]
	\ell(m_{Z}^{2})
	\nonumber \\&&{}
	+\biggl[R_{\eta}(m_{X},m_{X})+R_{\eta}(m_{X},m_{Z})
	+R_{\eta}(m_{Z},m_{Z})\biggr]\ell(m_{\eta}^{2})
	\nonumber \\&&{}
  	- 6\ell(m_{xd}^{2})-3\ell(m_{xs}^{2})-6\ell(m_{zd}^{2})
	-3\ell(m_{zs}^{2})
	\nonumber \\&&{}
	-R_{X}(m_{\eta})\tilde{\ell}(m_{X}^{2})-R_{Z}(m_{\eta})
	\tilde{\ell}(m_{Z}^{2})\biggr\}.
\label{eq:O81 Kvac logs2}
\end{eqnarray}
For $K\to\pi$, we again separate the logarithmic terms proportional to $\alpha_1$ and $\alpha_2$,
\begin{equation}
 	\left\langle \pi^+\left|\mathcal{O}^{\left(8,1\right)}
	\right|K^{+}\right\rangle_{\rm logs}
	=
 	\left\langle \pi^+\left|\mathcal{O}^{\left(8,1\right)}
	\right|K^{+}\right\rangle_{\rm logs}^{(1)}
	+
 	\left\langle \pi^+\left|\mathcal{O}^{\left(8,1\right)}
	\right|K^{+}\right\rangle_{\rm logs}^{(2)}\ ,
\end{equation}
with
\begin{eqnarray*}
 	\left\langle \pi^+\left|\mathcal{O}^{\left(8,1\right)}
	\right|K^{+}\right\rangle_{\rm logs}^{(1)} 
 	& = & \frac{4\alpha_{1}}{f^{2}}m_{xz}m_{X}
	\left[\frac{1}{2}\delta Z_{X}+\frac{1}{2}\delta Z_{xz}
	+\frac{1}{2}\left(\frac{\delta m_{X}^{2}}{m_{X}^{2}}
	+\frac{\delta m_{xz}^{2}}{m_{xz}^{2}}\right)\right]
	\\&&{}
	+ \frac{1}{9}\frac{\alpha_{1}}{f^{4}}\biggl\{
	36\biggl[
	 2m_{xd}^2 - m_X^2 - m_{xz}m_{X}
	+\frac{m_{X}(m_{zd}^{2}+m_{xd}^{2})}{m_{xz}-m_{X}}
	\biggr]\ell(m_{xd}^{2})
	\\&&{}
	+18\biggl[
	2 m_{xs}^2 - m_X^2
	-m_{xz}m_{X}
	+\frac{m_{X}(m_{zs}^{2} + m_{xs}^{2})}{m_{xz}-m_{X}}
	\biggr]\ell(m_{xs}^{2})
	\\&&{}  
	-12\biggl[
	6 m_{zd}^2 + 3m_X^2
	+m_{xz}m_{X}
	+\frac{3m_X(m_{zd}^{2} + m_{xd}^{2})}{m_{xz}-m_{X}}
	\biggr]\ell(m_{zd}^{2})
	\\&&{} 
	-6\biggl[
	6m_{zs}^2 + 3m_X^2 + m_{xz}m_{X} + \frac{3m_{X}(m_{zs}^{2}
	+m_{xs}^{2})}{m_{xz}-m_{X}}\biggr]\ell(m_{zs}^{2})
	\\&&{}
	+2\biggl[\biggl(-4m_{\eta}^{2}+4m_{xz}m_{X}-2m_{X}^{2}
	-\frac{3m_{X}(m_{\eta}^{2} + m_{X}^2)}{m_{xz}-m_{X}}\biggr)
	R_{\eta}(m_{X},m_{X})
	\\&&{}
	+2\biggl(2m_{xz}^{2}-4m_{xz}m_{X}+m_{X}^{2}\biggr)
	R_{\eta}(m_{X},m_{Z})
	\\&&{}
	+\biggl(4m_{\eta}^{2}+4m_{xz}^{2}+4m_{xz}m_{X}+4m_{X}^{2}
	\\&&{}
	+\frac{3m_{X}(m_{\eta}^{2} + m_X^2)}{m_{xz}-m_{X}}\biggr)
	R_{\eta}(m_{Z},m_{Z})\biggr]
	\ell(m_{\eta}^{2})\biggr\}
\end{eqnarray*}

\begin{eqnarray*}
	\qquad&&{}
	+ \frac{2}{9}\frac{\alpha_{1}}{f^{4}}\biggl\{\biggl[
	2m_X m_{xz}\frac{5m_X - 2m_{xz}}{m_{xz}-m_{X}}
	\left(R_{\eta}(m_{X},m_{X}) - 1\right)
	-\frac{4 m_{xz} - m_{X}}{m_{xz}-m_{X}}
	R_{X}(m_{\eta})
	\\&&{}
	+2\biggl(2m_{xz}^{2}-4m_{xz}m_{X}+m_{X}^{2}\biggr)
	R_{X}(m_{Z},m_{\eta})\biggr]\ell(m_{X}^{2})
	\\&&{}
	+\biggl[
	\frac{4m_{xz} - m_{X}}{m_{xz}-m_{X}}R_{Z}(m_{\eta})
	-\biggl(4m_{xz}(3m_{xz} + m_X)
	+\frac{6m_{X}m_{xz}^2}{m_{xz}-m_{X}}\biggr)
	\left(R_{\eta}(m_{Z},m_{Z}) - 1\right)
	\\&&{}
	+2\biggl(2m_{xz}^{2}-4m_{xz}m_{X}+m_{X}^{2}\biggr)
	R_{Z}(m_{X},m_{\eta})\biggr]\ell(m_{Z}^{2})\biggr\}
	\\&&{}
	+ \frac{2}{9}\frac{\alpha_{1}}{f^{4}}\biggl\{
	-12m_{xz} m_X \ell(m_{xz}^{2})
	+2m_{X}\biggl[3m_{X} - 2m_{xz}
	+\frac{3m_{X}^2}{m_{xz}-m_{X}}\biggr]
	R_{X}(m_{\eta})\tilde{\ell}(m_{X}^{2})
	\\&&{}
	-2m_{xz}\biggl[2(3m_{xz} + m_{X})
	+\frac{3m_{X}m_{xz}}{m_{xz}-m_{X}}\biggr]
	R_{Z}(m_{\eta})\tilde{\ell}(m_{Z}^{2})\biggr\}
	\\&&{}
	+ \frac{1}{9}\frac{\alpha_{1}}{f^{4}}\biggl\{
	36\biggl[
	m_X^2 (4m_{xz}^2  - m_{xd}^2 - m_{zd}^2)
	+(m_{xd}^2 + m_{zd}^2 - 2m_X^2 - 2m_{xz}^2)m_{xz}m_{X}
	\\&&{}
	+\frac{m_{X}(m_{zd}^4 - m_{xd}^4)}{m_{xz}-m_{X}}
	\biggr]\beta(q^{2},m_{zd}^{2},m_{xd}^{2})
	\\&&{}
	+18\biggl[
	m_X^2( 4m_{xz}^2 - m_{xs}^2 - m_{zs}^2)
	+(m_{xs}^2 + m_{zs}^2 - 2m_{xz}^2 - 2m_X^2)m_{xz}m_{X}
	\\&&{}
	+\frac{m_{X} (m_{zs}^{4} - m_{xs}^{4})}{m_{xz}-m_{X}}
	\biggr]\beta(q^{2},m_{zs}^{2},m_{xs}^{2})
	\\&&{}
	+2\biggl[\biggl(2m_{\eta}^{4}+4m_{\eta}^{2}m_{xz}m_{X}
	+2m_{\eta}^{2}m_{X}^{2}-4m_{xz}m_{X}^{3}-4m_{X}^{4}
	\\&&{}
	+\frac{3m_{X}(m_{\eta}^{4} - m_{X}^4)}{m_{xz}-m_{X}}
	\biggr)R_{\eta}(m_{X},m_{X})
	\\&&{}
	+2\biggl(2m_{\eta}^{2}m_{xz}^{2}+4m_{xz}^{3}m_{X}
	+m_{\eta}^{2}m_{X}^{2}-2m_{xz}^{2}m_{X}^{2}
	+2m_{xz}m_{X}^{3}-m_{X}^{4}\biggr)R_{\eta}(m_{X},m_{Z})
	\\&&{}
	+\biggl(-2m_{\eta}^{4}+4m_{\eta}^{2}m_{xz}^{2}
	-4m_{\eta}^{2}m_{xz}m_{X}+8m_{xz}^{3}m_{X}
	-4m_{xz}^{2}m_{X}^{2}+8m_{xz}m_{X}^{3}
	\\&&{}
	+2m_{X}^{4}+\frac{3m_{X}(m_{X}^4-m_{\eta}^{4})}{m_{xz}-m_{X}}
	\biggr)R_{\eta}(m_{Z},m_{Z})\biggr]
	\beta(q^{2},m_{\eta}^{2},m_{xz}^{2})
	\biggr\}
\end{eqnarray*}

\begin{eqnarray}
	\qquad&&{}
	+\frac{2}{9}\frac{\alpha_{1}}{f^{4}}\biggl\{
	2m_{X}m_{xz}\biggl[
	\frac{2m_{xz} + m_{X}}{m_{xz}-m_{X}}R_{X}(m_{\eta})
	\nonumber\\&&{}
	+2\biggl(2m_{xz}^{2}+m_{X}^2\biggr)
	R_{X}(m_{Z},m_{\eta})\biggr]\beta(q^{2},m_{xz}^{2},m_{X}^{2})
	\nonumber\\&&{}
	+\biggl[
	12m_X^3m_{xz}
	+\frac{3m_{X}(m_{X}^4-m_{Z}^{4})}{m_{xz}-m_{X}}
	+2\biggl(
	-2 m_{xz}^2 - 2 m_{xz}m_X + 2 m_X^2 
	\nonumber\\&&{}
	- \frac{3 m_Z^2 m_X}{m_{xz} - m_X}
	\biggr)R_{Z}(m_{\eta})
	\nonumber\\&&{}
	+2\biggl(
	m_Z^4-2m_X^4 + m_X^2 m_Z^2
	+2(2m_{xz}^2+m_{X}^2)m_X m_{xz}
	\biggr)R_{Z}(m_{X},m_{\eta})
	\nonumber\\&&{}
	+\biggl(
	-12 m_{xz}m_X^3 + \frac{3 m_X(m_Z^4-m_X^4)}{m_{xz}-m_X}\biggr)
	R_{\eta}(m_{Z},m_{Z})\biggr]
	\beta(q^{2},m_{Z}^{2},m_{xz}^{2})
	\nonumber\\&&{}
	+\biggl[
	4(2m_{X}^2 + 2m_{xz}^{2} - m_{Z}^{2})m_{xz}m_{X}
	\nonumber\\&&{}
	+\frac{3m_X(m_{X}^{4}-m_{Z}^{4})}{m_{xz}-m_{X}}\biggr]
	R_{Z}(m_{\eta})\tilde{\beta}(q^{2},m_{Z}^{2},m_{xz}^{2})
	\biggr\},\label{eq:O81 Kpi logs1}
\end{eqnarray}
and
\begin{align*}
	\left\langle \pi^{+}\left|\mathcal{O}^{\left(8,1\right)}
	\right|K^{+}\right\rangle_{\rm logs}^{(2)} =&	
	-\frac{4\alpha_{2}}{f^{2}}m_{xz}^{2}\left[
	\frac{1}{2}\delta Z_{X}+\frac{1}{2}\delta Z_{xz}\right]
	\nonumber\\
	 & + \frac{4}{9}\frac{\alpha_2}{f ^{4} }m_{xz}^{2}
	 \biggl\{-\frac{9m_{X}(2\ell(m_{xd}^{2})  
	 +\ell(m_{xs}^{2}))}{m_{xz}-m_{X}} \nonumber\\	 
 	& +\biggl[-2  +\frac{3m_{X}}{m_{xz}-m_{X}}  
	+\biggl(2  -\frac{3m_{X}}{m_{xz}-m_{X}}\biggr)R_{\eta}(m_{X},m_{X})  
	\nonumber\\&
	-2R_{X}(m_{Z},m_{\eta})\biggr]\ell(m_{X}^{2}) \nonumber\\
 	& +\biggl[-2  -\frac{3m_{X}}{m_{xz}-m_{X}}  
	+\biggl(2  +\frac{3m_{X}}{m_{xz}-m_{X}}\biggr)R_{\eta}(m_{Z},m_{Z})  
	\nonumber\\&
	-2R_{Z}(m_{X},m_{\eta})\biggr]\ell(m_{Z}^{2}) \nonumber\\
 	& +6\biggl[2  +\frac{3m_{X}}{m_{xz}-m_{X}}\biggr]\ell(m_{zd}^{2})  
	+3\biggl[2  +\frac{3m_{X}}{m_{xz}-m_{X}}\biggr]\ell(m_{zs}^{2}) \nonumber\\
 	& +\biggl[\biggl(-2  +\frac{3m_{X}}{m_{xz}-m_{X}}\biggr)R_{\eta}(m_{X},m_{X})  
	-2R_{\eta}(m_{X},m_{Z})  \nonumber\\
	& -\biggl(2  +\frac{3m_{X}}{m_{xz}-m_{X}}\biggr)R_{\eta}(m_{Z},m_{Z})\biggr]
	\ell(m_{\eta}^{2})
\end{align*}

\begin{align}
 & -\biggl[\biggl(2  +\frac{3m_{X}}{m_{xz}-m_{X}}\biggr)R_{X}(m_{\eta})  
 +2\biggl(2m_{xz}^{2}  +m_{X}^{2}\biggr)R_{X}(m_{Z},m_{\eta})\biggr]
 \beta(q^{2},m_{xz}^{2},m_{X}^{2}) \nonumber\\
 & +\biggl[2m_{Z}^{2}  -4m_{xz}^{2}  
 +\frac{3m_{X}(m_{Z}^{2}  -m_{X}^{2})}{m_{xz}-m_{X}}  -4m_{X}^{2}  
 +\biggl(2  +\frac{3m_{X}}{m_{xz}-m_{X}}\biggr)R_{Z}(m_{\eta}) \nonumber\\
 & +\biggl(-2m_{Z}^{2}  +4m_{xz}^{2}  +\frac{3m_{X}(-m_{Z}^{2}  
 +m_{X}^{2})}{m_{xz}-m_{X}}  +4m_{X}^{2}\biggr)R_{\eta}(m_{Z},m_{Z}) \nonumber\\
 & -2\biggl(2m_{xz}^{2}  +m_{X}^{2}\biggr)R_{Z}(m_{X},m_{\eta})\biggr]
 \beta(q^{2},m_{Z}^{2},m_{xz}^{2}) \nonumber\\
 & +6\biggl[-2m_{zd}^{2}  +2m_{xd}^{2}  +2m_{xz}^{2}  
 +3m_{X}\biggl(\frac{-m_{zd}^{2}  +m_{xd}^{2}}{m_{xz}-m_{X}}  -m_{xz}\biggr)
 +m_{X}^{2}\biggr]\beta(q^{2},m_{zd}^{2},m_{xd}^{2}) \nonumber\\
 & +3\biggl[-2m_{zs}^{2}  +2m_{xs}^{2}  +2m_{xz}^{2}  
 +3m_{X}\biggl(\frac{-m_{zs}^{2}  +m_{xs}^{2}}{m_{xz}-m_{X}}  -m_{xz}\biggr)
 +m_{X}^{2}\biggr]\beta(q^{2},m_{zs}^{2},m_{xs}^{2}) \nonumber\\
 & +\biggl[\biggl(-2m_{\eta}^{2}  +\frac{3m_{X}(-m_{\eta}^{2}  +m_{X}^{2})}
 {m_{xz}-m_{X}}  +2m_{X}^{2}\biggr)R_{\eta}(m_{X},m_{X}) \nonumber\\
 & -2\biggl(2m_{xz}^{2}  +m_{X}^{2}\biggr)R_{\eta}(m_{X},m_{Z}) \nonumber\\
 & +\biggl(2m_{\eta}^{2}  -4m_{xz}^{2}  +\frac{3m_{X}(m_{\eta}^{2}  
 -m_{X}^{2})}{m_{xz}-m_{X}}  -4m_{X}^{2}\biggr)R_{\eta}(m_{Z},m_{Z})\biggr]
 \beta(q^{2},m_{\eta}^{2},m_{xz}^{2})    \nonumber\\
 & + \biggl[2  -\frac{3m_{X}}{m_{xz}-m_{X}}\biggr]R_{X}(m_{\eta})\tilde{\ell}(m_{X}^{2})
 +\biggl[2  +\frac{3m_{X}}{m_{xz}-m_{X}}\biggr]R_{Z}(m_{\eta})\tilde{\ell}(m_{Z}^{2}) \nonumber\\
 & + \biggl[2m_{Z}^{2}  -4m_{xz}^{2}  +\frac{3m_{X}(m_{Z}^{2}  -m_{X}^{2})}{m_{xz}-m_{X}}
 -4m_{X}^{2}\biggr]
 R_{Z}(m_{\eta})\tilde{\beta}(q^{2},m_{Z}^{2},m_{xz}^{2}) \biggr\}. 	\label{eq:O81 Kpi logs2}
\end{align}

These formulas are simplified enormously in the degenerate valence case,
\begin{eqnarray}
	 \left\langle \pi^{+}\left|\mathcal{O}^{\left(8,1\right)}
	 \right|K^{+}\right\rangle ^{\rm deg.val.,(1)}_{\rm logs}
	 & = & 
	\frac{4\alpha_{1}}{f^{2}}m_{X}^{2}\left[
	\delta Z_{X}+\frac{\delta m_{X}^{2}}{m_{X}^{2}}\right]
	\nonumber \\&&{}
	+ \frac{4}{3}\frac{\alpha_{1}}{f^{4}}m_{X}^{2}
	\biggl\{
	2\biggl[\ell(m_{\eta}^{2})
	\nonumber \\&&{}
	+\biggl(m_{\eta}^{2} + m_{X}^2\biggr)
	\beta(0,m_{\eta}^{2},m_{X}^{2})\biggr]
	R_{\eta}(m_{X},m_{X})
	\nonumber \\&&{}
 	+2 \biggl[2 - R_{\eta}(m_{X},m_{X})\biggr]
	\ell(m_{X}^{2})
	-10\ell(m_{xd}^{2})
	\nonumber \\&&{}
 	 -5 \ell(m_{xs}^{2})
 	-2m_{X}^2R_{X}(m_{\eta})\tilde{\tilde{\ell}}(m_{X}^{2})
	\nonumber \\&&{}
 	+4\biggl[-m_{X}^2 - R_{X}(m_{\eta})
	+ m_X^2R_{\eta}(m_{X},m_{X})\biggr]
	\tilde{\ell}(m_{X}^{2})	\biggr\},
	\label{eq:O81 Kpi deg val logs1}
\end{eqnarray}
\begin{align}
 	\left\langle \pi^{+}\left|\mathcal{O}^{\left(8,1\right)}
	 \right|K^{+}\right\rangle ^{\rm deg.val.,(2)}_{\rm logs}
	 = & 
	-\frac{4\alpha_{2}}{f^{2}}m_{X}^{2}\delta Z_{X}
	\nonumber \\
	& + \frac{8}{3}\frac{\alpha_2}{f ^{4} }m_{X}^{2}
	 \biggl\{m_{X}^{2}\biggl[R_{X}(m_{\eta})\tilde{\tilde{\ell}}(m_{X}^{2})
	 -2R_{\eta}(m_{X},m_{X})\beta(0,m_{\eta}^{2},m_{X}^{2})\biggr] \nonumber\\
 	&  +\biggl[-1  +R_{\eta}(m_{X},m_{X})\biggr]\ell(m_{X}^{2})  
	+2\ell(m_{xd}^{2})  +\ell(m_{xs}^{2})  \nonumber\\
	& -R_{\eta}(m_{X},m_{X})\ell(m_{\eta}^{2}) \nonumber\\
 	&  +\biggl[2m_{X}^{2}\biggl(1  -R_{\eta}(m_{X},m_{X})\biggr)
	+R_{X}(m_{\eta})\biggr]\tilde{\ell}(m_{X}^{2})\biggr\}.
					\label{eq:O81 Kpi deg val logs2}
\end{align}

\section{Erratum \label{sub:erratum}}

We note here some corrections to the works of Refs.~\cite{Laiho:2002jq} and \cite{Laiho:2003uy}.  All of the NLO low energy constants for the (27,1) operators have the wrong sign in both Ref.~\cite{Laiho:2002jq} and Ref.~\cite{Laiho:2003uy}.  The values for $\gamma_i$ appearing in Table I of Ref.~\cite{Laiho:2002jq} (and again in Table I of Ref.~\cite{Laiho:2003uy}) should have the opposite sign.  In Eq.~(16) of Ref.~\cite{Laiho:2002jq} (and again in Eq.~(16) of Ref.~\cite{Laiho:2003uy}), the operators ${\cal O}^{(8,1)}_5$ and ${\cal O}^{(8,1)}_{15}$ should have opposite sign to be consistent with the signs of the LEC's $e^r_5$ and $e^r_{15}$ appearing in the amplitudes presented in those works. We make these corrections in the current work.  Since the LEC's are not known, an incorrect, but consistent normalization of them (including an incorrect sign) does not alter the procedure of using the formulas of Ref.~\cite{Laiho:2003uy} to construct $K\to\pi\pi$ from $K\to\pi$ and $K\to 0$ matrix elements.  Therefore, these corrections make no difference to the conclusions of these works that it is possible to obtain all of the LEC's needed to construct $K\to\pi\pi$ matrix elements through NLO in $\chi$PT from lattice calculations.  

	Additionally in Ref.~\cite{Laiho:2003uy}, there is a typo in the (8,8) $K\to\pi\pi$ matrix element formulas.  The coefficient of the second term in Eq.~(42) should be $\frac{12i}{f_K f_\pi^2}$, and the coefficient of the second term in Eq.~(43) should be $-\frac{12i}{f_K f_\pi^2}$.  The corrected versions of these equations are given in the current work as Eqs.~(\ref{eq:K2pipi883/2}) and (\ref{eq:K2pipi881/2}).  These corrections also do not alter the conclusions of Ref.~\cite{Laiho:2003uy}, but are necessary to construct the correct $K\to\pi\pi$ matrix elements.
 
\bibliographystyle{apsrev}
\bibliography{samref}

\end{document}